\documentclass{jfm}
\usepackage{graphicx}
\usepackage{xcolor}
\usepackage{subcaption}
\usepackage{placeins}
\usepackage[colorlinks=true,citecolor=blue,linkcolor=blue,filecolor=blue]{hyperref}

\title{Dewetting of a corner film wrapping a wall-mounted cylinder}

\author{Si Suo\aff{1}, Seyed Morteza Habibi Khorasani\aff{1} \and Shervin Bagheri\aff{1} \corresp{\email{shervin@mech.kth.se}}
 }

\shortauthor{S. Suo, S. M. Habibi Khorasani and S. Bagheri}

\affiliation{\aff{1}FLOW, Department of Engineering Mechanics, KTH Royal Institute of Technology, SE-100 44 Stockholm, Sweden
}

\begin{document}

\maketitle

\begin{abstract}
In this study, we investigate the stability of a liquid film that partially wets a corner between a cylinder and a substrate, using a combination of theoretical and numerical approaches. The film stability, which depends on the film size and the wall wettability, is firstly predicted by a standard linear stability analysis (LSA) within the long-wave theoretical framework. We find that the film size plays the most important role in controlling the film stability. Specifically, the thinner the film is, the more sensitive it becomes to the large-wavenumber perturbation. The wall wettability mainly impacts the growth rates of perturbations and slightly influences the marginal stability and post\nobreakdash-instability patterns of wrapping films. We compare the LSA predictions with numerical results obtained from a disjoining pressure model (DPM) and Volume-of-Fluid (VOF) simulations, which provide more insights into the film breakup process. At the early stage there is a strong agreement between the LSA predictions and the DPM results. Notably, as the perturbation grows, thin film regions connecting two neighboring satellite droplets form which may eventually lead to a stable or temporary secondary droplet, an aspect which the LSA is incapable of capturing. In addition, the VOF simulations suggest that beyond a critical film size the crest coalescence mechanism becomes involved during the breakup stage. Therefore, the LSA predictions are able to provide only an upper limit on the final number of satellite droplets.
\end{abstract}


\section{Introduction}
The stability of liquid filaments, threads, or films is an important branch of fluid dynamics. The pioneering research dates back to the discovery of the Plateau-Rayleigh instability, which describes the mechanism governing the breakup of falling fluid streams into smaller droplets due to their surface tension driven tendency of minimizing their surface area \citep{plateau1850ueber}. The instability problem becomes more challenging when considering fluid-solid interactions, such as a liquid filament lying upon a solid substrate \citep{benilov2009stability} or within a groove \citep{sundin2021roughness}, and film dewetting on patterned surfaces \citep{kim2018viscous,martin2021prediction}, where complex interfacial dynamics involving topological breakup, coalescence, etc. are present. An understanding of the relevant stability conditions and final interfacial morphology can benefit a wide range of engineering applications, especially in the field of microfluidics \citep{olanrewaju2018capillary,chen2019confinements,yasuga2021fluid}. Therefore, it is necessary to understand the effects influencing interfacial evolution and breakup patterns.

The simplest situation to consider would be a liquid filament or rivulet partially wetting a substrate or a groove, for which the related stability problem has been extensively studied. In an early study, \cite{langbein1990shape} theoretically investigated the stability of a liquid filament attached to solid edges at an arbitrary angle using quasi-static solutions of the Laplace equation and geometric constraints. He established a critical filament length beyond which the filament becomes unstable.
Similarly based on the quasi-static assumption, \cite{roy1999stability} derived a more general stability criterion where the filament stability is guaranteed if the capillary pressure is an increasing function of the filament cross-sectional area. In other words, the stability criterion derived from quasi-static solutions is also a measure of whether it is energetically favourable for a film to breakup or not \citep{wilson2005energetically}.
Assuming a shallow liquid film and a small Reynolds number for the flow, the lubrication approximation has also been widely adopted to derive time-dependent equations for film thickness evolution \citep{hocking1990spreading,hocking1993stability}. It has been shown that the lubrication approximation yields reasonable predictions for film flows compared to Navier-Stokes solutions \citep{perazzo2004navier}. This approximation allows for the consideration of gravity and moving contact lines, enabling the modeling of more complex interfacial dynamics.
For instance, considering contact angle hysteresis, researchers have thoroughly investigated the competing effects between gravity and surface tension on the stability of an infinite rivulet interacting with an inclined plane \citep{hocking1993stability,benilov2009stability}. Interestingly, \cite{benilov2009stability} pointed out that if the rivulet is on the underside of the inclined plane where the gravity effect is inverted, gravity-induced shear flows could act as a stabilizing factor, contrary to the typical perception of it being a source of instability \citep{hocking1993stability,myers2004stability,sullivan2008thin}.
Furthermore, numerical frameworks based on the disjoining pressure model (DPM) have been developed for filaments of finite length \citep{diez2007breakup,diez2009breakup}. The results from using this framework uncovered two dewetting patterns; either the filament shrinks and eventually forms a single droplet if it is short and thick, or it breaks up into several sub-droplets. Importantly, a close connection to the Plateau-Rayleigh instability was highlighted, suggesting the same underlying physical mechanisms. Studies employing similar theoretical and numerical techniques but involving other configurations, such as a liquid ring on a solid substrate \citep{gonzalez2013stability,edwards2021controlling} and irregular liquid structures \citep{huang2017stability}, have also followed.

At present, more research is focused on liquid structures interacting with confined geometries, such as tubes and corners, owing to their ability to guide a specific fluid phase in a controllable manner, especially in the fields of artificial surfaces and microfluidics. Studies on the stability of liquid structures trapped within tubes and corners have been reported throughout the literature \citep{chen2006capillary,yang2007capillary,lv2021wetting}. However, a deeper understanding of many aspects of such configurations is warranted.

\section{Cylinder-wrapping corner film}
The \emph{cylinder-plane-corner} geometry is the focus of this work as cylinders are commonly used or encountered in artificial porous media or microfluidic devices \citep{holtzman2015wettability,zhao2016wettability,rabbani2018suppressing,suo2022mobility}. The fluid dynamics in the corner regions of such geometries plays an essential role in flow patterning. Relevant work has mainly been focused on the process of film formation, i.e. the imbibition along the corners. This so-called corner flow, which emerges as a specific pattern of fluid-fluid displacement in porous media \citep{primkulov2021wettability}, occurs under strong imbibition \citep{zhao2016wettability,hu2018wettability} and is also known as the Concus–Finn condition \citep{concus1969behavior}. However, the dewetting process has seldom been given attention to. Specifically, once beyond the Concus–Finn condition, capillary-induced instability may occur in the formed wrapping films.

\begin{figure}
\centerline{\includegraphics[width=0.9\linewidth]{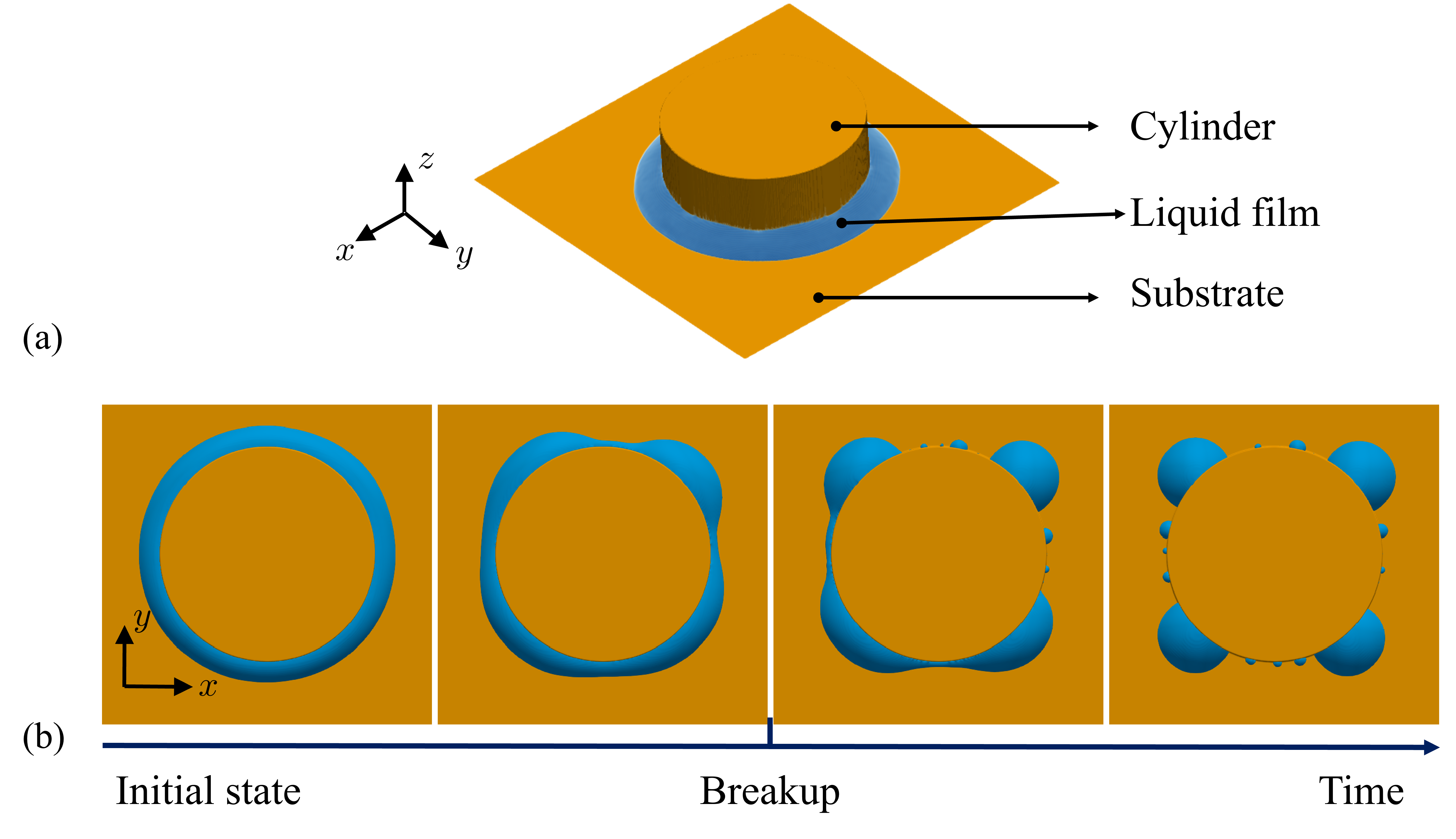}}
  \captionsetup{justification=justified}
  \caption{(a) Numerical realization of a liquid film wrapping a cylinder. (b). Top views of the morphology evolution from the initial state to final breakup. The ratio of film size on the cylinder radius is 0.22; the contact angle on the cylinder and the substrate is $90^{\circ}$ and $30^{\circ}$, respectively.}
\label{fig:f0}
\end{figure}

\begin{figure}
\centerline{\includegraphics[width=0.9\linewidth]{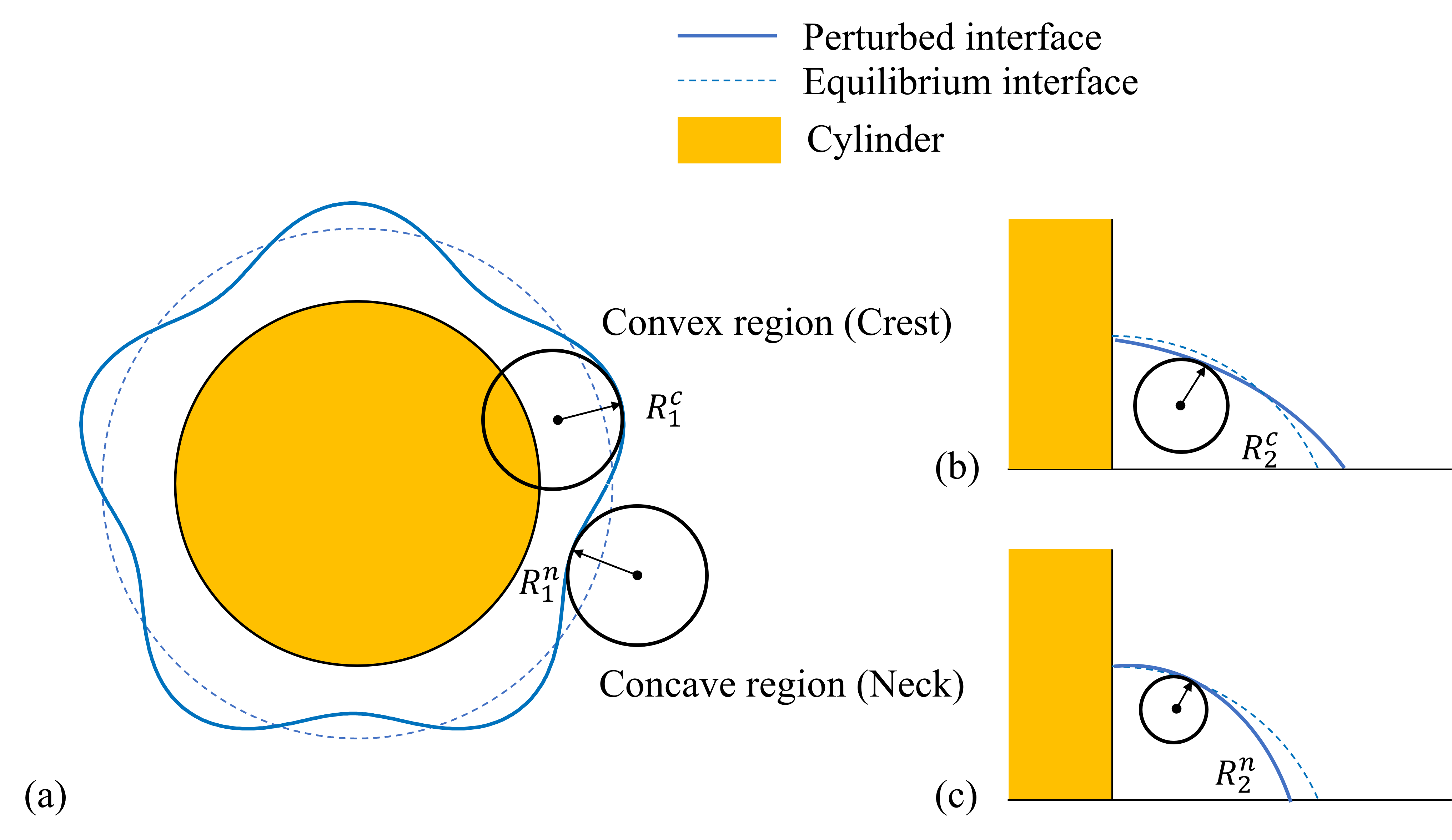}}
  \captionsetup{justification=justified}
  \caption{Schematic of a perturbed corner film from two principal directions. (a) is the top view; (b) and (c) are side views corresponding to the convex and concave region, respectively. }
\label{fig:f0.1}
\end{figure}

A film-cylinder system is presented in figure \ref{fig:f0}(a). It shows a settled film on a flat surface at a contact angle of $30^{\circ}$ which is also wrapped around a cylinder at a contact angle of $90^{\circ}$. An example of how the breakup of such a film evolves --obtained using a volume-of-fluid (VOF) simulation-- is shown in figure \ref{fig:f0}(b). One can see how the perturbation in the film grows over time and finally causes it to break up into four satellite droplets. Figure \ref{fig:f0.1} depicts a conceptual examination of this instability mechanism. The perturbation is assumed as a superposition of sinusoidal components with various wavenumbers. Considering only a single component where the wavenumber is five, in figure \ref{fig:f0.1}(a) the perturbed interface shows two types of regions which are convex and concave with respect to the equilibrium state. At the convex region (crest), the curvature ($1/R^c_1$) is positive, yielding a larger capillary pressure compared to the one at the concave region (neck) where the curvature ($1/R^n_1$) is negative. Consequently, the liquid tends to be squeezed from the convex to the concave region, eventually returning the film to its equilibrium state. However, the capillary pressure, as determined by the Young-Laplace equation, also depends on the curvature in the other principal direction, as shown in figures \ref{fig:f0.1}(b) and (c). The curvature ($1/R^c_2$) in the convex region may become smaller while the curvature ($1/R^n_2$) in the concave region may become larger due to the perturbation. Thus, the latter effect can counteract the former and prevent the liquid from flowing from the convex to the concave regions. These two effects are determined by the wavenumber, wettability conditions and film size. If they cannot balance each other out, the perturbed component would grow or decay over time, corresponding to a positive or negative growth rate. Generally, the component with the maximum growth rate would dominate the instability process and eventually pinch the film into satellite droplets, the number of which can be predicted by the corresponding wavenumber as shown in figure \ref{fig:f0}(b).


Modeling the development of this instability and predicting the consequent morphology patterns would be very beneficial. As such, our work focuses on two questions; first, under what conditions does the corner film become unstable, and second, how many satellite droplets are formed once the instability occurs. The paper is organized as follows. In section \ref{LSA}, the long-wave theory scheme is used to formulate the governing equation of the film thickness and introduce the corresponding boundary conditions. The equilibrium profile is solved for a certain corner condition, based on which we perform a standard linear stability analysis (LSA) to obtain the growth rate. The dependence of growth rate on the corner conditions is then thoroughly investigated. To gain more understanding of interfacial dynamics during the film breakup process and validate the LSA predictions, a numerical scheme using the disjoining pressure model (DPM) for directly tracking the evolution of film morphology is developed in section \ref{sec:DPM}. Additionally, numerical simulations involving solutions of the Navier-Stokes equations for two-fluid systems using the volume-of-fluid (VOF) method, which abandon the assumptions used in LSA and DPM, are conducted in section \ref{sec:VOF}. We ultimately demonstrate to what extent the simplified theoretical model can explain the phenomena observed in the VOF simulations and explain the applicability of the LSA predictions.

\section{Linear stability analysis (LSA)} \label{LSA}
\begin{figure}
  \centerline{\includegraphics[width=0.9\linewidth]{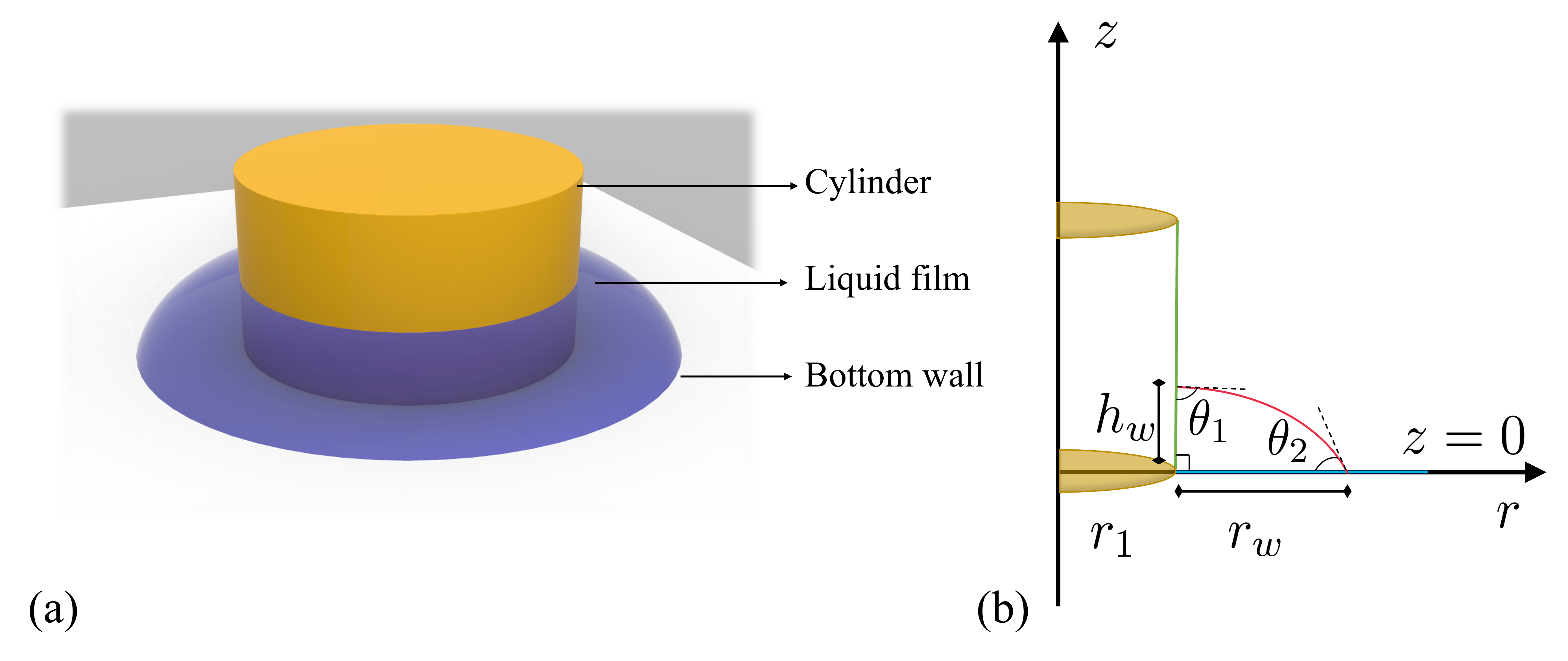}}
  \captionsetup{justification=justified}
  \caption{(a) Illustration of a corner liquid film around a vertical cylinder along with (b) a schematic describing the relevant geometric parameters.}
\label{fig:f1}
\end{figure}

We first develop a mathematical model for describing the evolution of a corner film, obtain its equilibrium state and finally perform linear stability analysis. Figure \ref{fig:f1} shows a schematic of the problem under consideration, a liquid film wrapping a vertical cylinder, along with the relevant geometric parameters including the cylinder radius $r_1$, wetting radius $r_w$ and height $h_w$, contact angle on the cylinder $\theta_1$ and on the bottom wall $\theta_2$.

\subsection{Governing equations}
Limiting the configuration to $r_w \approx h_w$ and $r_w\ll r_1$, we can describe the flow within the long-wave theory framework ---also known as the lubrication assumption--- for a thin film with a free surface. Although this approach assumes small slopes everywhere ($\theta_1 \to 90^{\circ}$ and $\theta_2 \to 0^{\circ}$) comparison with numerical simulations indicates that the theoretical solution can provide reasonable results for slope angles smaller than $45^{\circ}$ \citep{perazzo2004navier,mahady2013comparison}. Therefore, we limit the parameter space to $\theta_1\in \left[75^{\circ}, \ 90^{\circ}\right]$, $\theta_2\in \left[15^{\circ}, \ 45^{\circ}\right]$ and $r_w/r_1\in\left[0.12,\ 0.30\right]$. Additionally, we assume that the characteristic size of the liquid film is smaller than the capillary length $l_c=\sqrt{\gamma/\rho g}$, where $\rho g$ is the specific weight of the liquid and $\gamma$ is the surface tension, so that gravity effects can be neglected. Combined with the Navier-slip boundary condition for addressing the so-called ``contact-line singularity'' \citep{shikhmurzaev2006singularities, savva2011dynamics}, the governing equation regarding the film thickness $h$ is built in a cylindrical coordinate system ($r$, $\phi$, $z$). Non-dimensionalizing the lengths with $r_1$ and time with $6\mu r_{1}/{\gamma}$, where $\mu$ is the liquid viscosity, gives \citep{hocking1993stability}
\begin{equation}\label{eqn:heqn}
\frac{\partial h}{\partial t}+\nabla \cdot\left[h^{2}(h+\ell) \nabla \nabla^{2} h\right]=0,
\end{equation}
where $\ell$ is the prescribed slip length and set as $10^{-3}$ here. The corresponding boundary conditions for the cylinder and bottom wall contact line can be expressed as
\begin{subequations} \label{eqn:hbc}
\begin{align}
h\left(r_1, t\right)=h_w\left(\phi, t \right),\\
h\left(r_1+r_w \left(\phi, t\right), t \right)=0,
\end{align}
\end{subequations}
and the contact angle is imposed as 
\begin{subequations}\label{eqn:thetabc}
\begin{align}
\left .\frac{\partial h}{\partial r} \right |_{r=r_1} &=-\cot\theta_1,\\
\left .\frac{\partial h}{\partial r} \right |_{r=r_w+r_1} &=-\tan\theta_2.
\end{align}\end{subequations}
The solution of \eqref{eqn:heqn}-\eqref{eqn:thetabc} is assumed to be a superposition of an equilibrium solution $h_{0}(r, \phi)$ and a perturbation $h_{1}(r, \phi, t)$,
\begin{equation}\label{eqn:h1h0}
h\left(r, \phi, t\right)=h_{0}\left(r, \phi \right)+\epsilon h_{1}\left(r, \phi, t\right),\\
\end{equation}
where $\epsilon \ll 1$. Substituting \eqref{eqn:h1h0} into \eqref{eqn:heqn}, we obtain the static equation for $h_0$,
\begin{equation}\label{eqn:steadyeqn}
\nabla^2 h_0=-p,
\end{equation}
where $p$ is a constant representing the capillary-induced pressure difference from the surrounding gas pressure. Neglecting the higher-order term $O\left(\epsilon^2 \right)$, the perturbation equation becomes
\begin{equation}\label{eqn:perturbationeqn}
\frac{\partial h_1}{\partial t}+\nabla \cdot\left[h_{0}^{2}\left(h_{0}+\ell\right) \nabla\left(\nabla^{2} h_{1}\right)\right]=0.
\end{equation}

\subsection{Equilibrium solution}

\begin{figure}
  \centerline{\includegraphics[width=13cm]{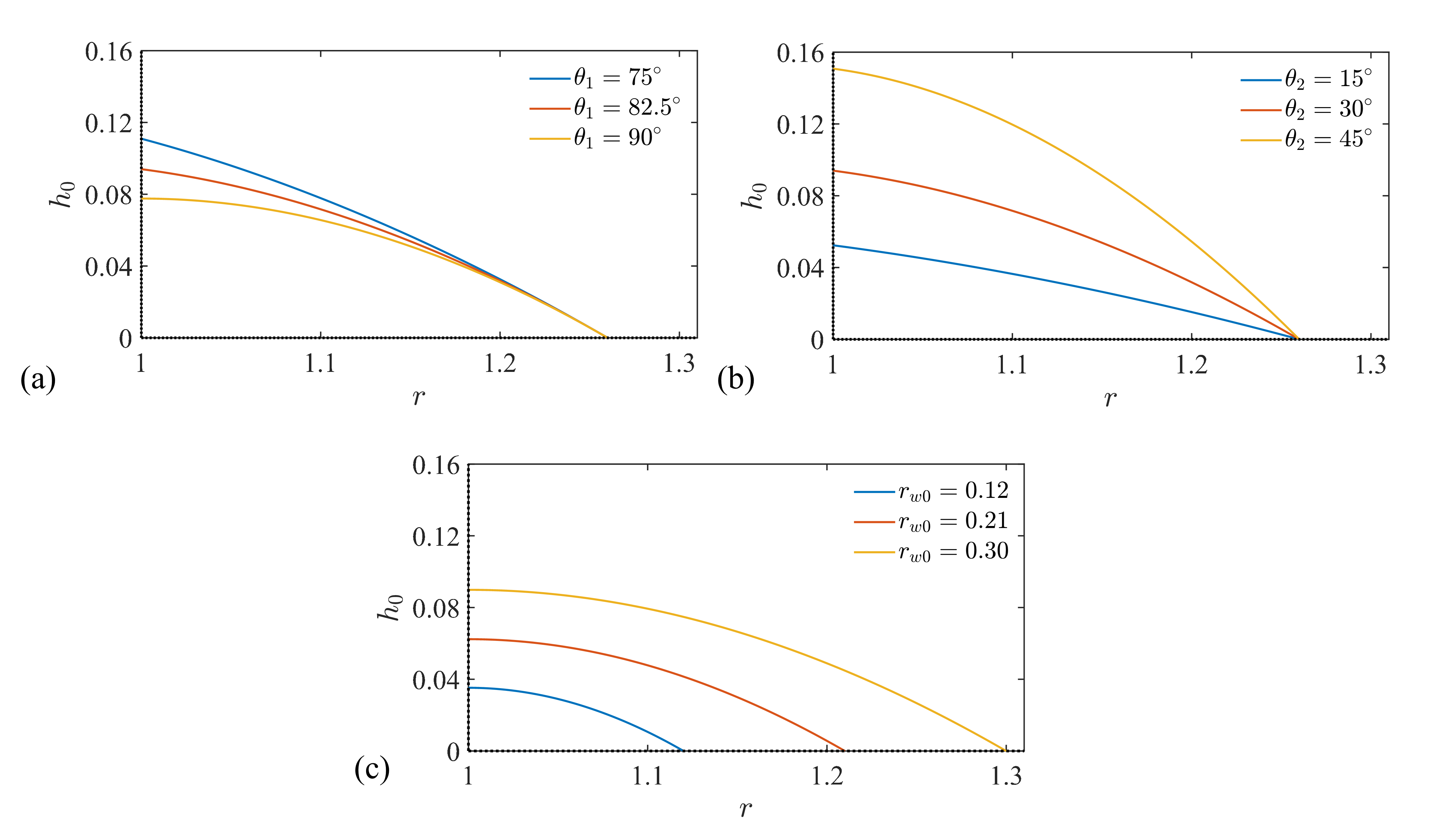}}
  \caption{Equilibrium interfacial profiles for (a) $r_{w0}=0.26$, $\theta_2 = 30^{\circ}$, and $\theta_1 = 75^{\circ},\ 82.5^{\circ},\ 90^{\circ}$; (b) $r_{w0}=0.26$, $\theta_1 = 82.5^{\circ}$, and $\theta_2 = 15^{\circ},\ 30^{\circ},\ 45^{\circ}$; and (c) $\theta_1 = 90^{\circ}$, $\theta_2 = 30^{\circ}$, and $r_{w0}=0.12,\ 0.21,\ 0.30$. }
\label{fig:h0}
\end{figure}

Considering an axisymmetric solution to \eqref{eqn:steadyeqn} gives
\begin{equation}\label{eqn:steadysol}
h_{0}(r)=-\frac{p r^{2}}{4}+c_{1} \ln r+c_{2},
\end{equation}
where the unknown parameters $c_1$, $c_2$, $p$ and $h_{w0}$ are determined by the boundary conditions \eqref{eqn:hbc} and \eqref{eqn:thetabc}, assuming $r_{w0}$ is given. The parameters are given by
\begin{subequations} \label{eqn:steadysol_paras}
\begin{align}
c_1 &= \frac{r_2}{r_2^2 - 1} \left( \tan \theta_2 - r_2 \cot \theta_1 \right),\\
c_2 &= \frac{r_2}{2\left(r_2^2 - 1 \right)} \left[2\ln r_2\left( r_2\cot \theta_1 -  \tan \theta_2\right) + r_2\left(r_2 \tan \theta_2 -  \cot \theta_1\right) \right],\\
p &= 2 \frac{r_2 \tan\theta_2 -  \cot \theta_1}{\left(r_2^2 - 1\right)},\\
h_{w0} &= c_{2}-\frac{1}{4} p,
\end{align}
\end{subequations}
where $r_2 = r_1 + r_{w0}$. The wettability condition ($\theta_1$ and $\theta_2$) and initial film size ($r_{w0}$), compose the corner condition, which completely determines the equilibrium interfacial shape. Figure \ref{fig:h0} compares the equilibrium profiles for different $\theta_1$, $\theta_2$, and $r_{w0}$. With a fixed $r_{w0}$, $\theta_1$ and $\theta_2$ directly change the interfacial curvature. As shown in figures \hyperref[fig:h0]{\ref*{fig:h0}(a)} and \hyperref[fig:h0]{\ref*{fig:h0}(b)}, the smaller $\theta_1$ and $\theta_2$ are (the more hydrophilic the solid walls are) the smaller the curvature of the meniscus becomes. When the wettability condition is fixed, figure \hyperref[fig:h0]{\ref*{fig:h0}(c)} shows that with increasing $r_{w0}$ the changes in $h_{w0}$ are relatively smaller, resulting in flatter corner films.

\subsection{Eigenvalue problem} \label{LSA_straight} 
The LSA is conducted using an equilibrium solution for a certain corner condition. Since the circumferential length of the film is much larger than its length along the $r$ and $z$ axes ($r_w \approx h_w\ll 2\pi r_1$), we can assume the perturbation is a periodic wave along the $\phi$ axis while the wave along the $r$ axis can be ignored. The perturbation $h_1$ thus takes the following form,
\begin{equation}\label{eqn:h1}
h_{1}\left(r, \phi, t\right)=\hat{h}_{1}(r) e^{\mathrm{i} \left(n \phi - \lambda t \right)}.
\end{equation}
In \eqref{eqn:h1}, $n$ is the angular wavenumber which should be small according to the long-wave assumption so $n\leq10$. Importantly, though $n$ is real, only the integer values of $n$ are physically meaningful as they correspond to the number of satellite droplets. Also, $\lambda = \mathrm{i} \sigma + \omega$ is the complex frequency composed of the growth rate $\sigma$ and the phase speed $\omega$. Consequently, the perturbation at the boundaries has the following form,
\begin{subequations}\label{eqn:h1BC}
\begin{align}
h_w(\phi, t) &= h_{w0}+\epsilon\xi_1e^{\mathrm{i} (n \phi - \lambda t)},\\
r_w(\phi, t) &= r_{w0}+\epsilon\xi_2e^{\mathrm{i} (n \phi - \lambda t)},
\end{align}
\end{subequations}
where $\xi_1$ and $\xi_2$ are two coefficients determined by the boundary conditions. Substituting \eqref{eqn:h1} into \eqref{eqn:perturbationeqn}, we obtain the eigenvalue problem,
\begin{equation}\label{eqn:L1h1}
\mathcal{L}(\hat{h}_1)=\mathrm{i}\lambda \hat{h}_1,
\end{equation}where 
\begin{equation} \label{eqn:L1}
\mathcal{L} \left(\hat{h}_1 \right) = \hat{c}_4(r, n) \hat{h}_{1, rrrr} +  \hat{c}_3(r, n) \hat{h}_{1, rrr} + \hat{c}_2(r, n) \hat{h}_{1, rr}+ \hat{c}_1(r, n) \hat{h}_{1, r} + \hat{c}_0(r, n) \hat{h}_{1}.
\end{equation}
The coefficients $\hat{c}_0$-$\hat{c}_4$ can be explicitly expressed as 
\begin{subequations}
\begin{align}
\hat{c}_4 &= H_0,\\
\hat{c}_3 &= 2 H_0 /r+H_{0r},\\
\hat{c}_2 &= H_{0r} /r -(2n^2+1)H_0/r^2,\\
\hat{c}_1 &= (2n^2+1)H_0 /r^3 - (n^2+2)H_{0r} /r^2+H_{0r} (h_{0rr}/r+h_{0rrr})/ h_{0r},\\
\hat{c}_0 &= H_{0r}h_{0rrrr}+(r H_1 h_{0r}+2H_{0r}/h_{0r})h_{0rrr}/r\\
&+(r H_1 h_{0r}-H_{0r}/h_{0r})h_{0rr}/r^2 - 2H_1(h_{0r}/r)^2\\
&+(2n^2+1)h_{0} H_{0r}/r^3+n^2(n^2-4)h^2_{0}(h_{0}+1)/r^4,
\end{align}
\end{subequations}
where $H_0=h^2_0(h_0+\ell$) and $H_1=2(3h_0+\ell)$.
The boundary conditions on $\hat{h}_1$ can be implemented by substituting \eqref{eqn:h1}-\eqref{eqn:h1BC} into \eqref{eqn:hbc}-\eqref{eqn:thetabc} and only keeping the linear terms. Specifically, on the cylinder wall,  
\begin{subequations}\label{eqn:h1BC_lin_left}
\begin{align}
h_0(r_1)+\epsilon\hat{h}_1(r_1)e^{\mathrm{i} (n \phi - \lambda t)}&=h_{w0}+\epsilon\xi_1e^{\mathrm{i} (n \phi - \lambda t)},\\
h_0'(r_1)+\hat{h}_1'(r_1)e^{\mathrm{i} (n \phi - \lambda t)}&=-\cot \theta_1; 
\end{align}
\end{subequations}
and on the bottom wall, 
\begin{subequations}\label{eqn:h1BC_lin_right}
\begin{align}
 h_0(r_2) + \epsilon\xi_2 h'_0(r_2)e^{\mathrm{i} (n \phi - \lambda t)} + \epsilon\hat{h}_1(r_1+r_w)e^{\mathrm{i} (n \phi - \lambda t)}&=0, \\
 h_0'(r_2) + \epsilon\xi_2 h''_0(r_2)e^{\mathrm{i} (n \phi - \lambda t)}+ \epsilon\hat{h}'_1(r_1+r_w)e^{\mathrm{i} (n \phi - \lambda t)}&=-\tan\theta_2. 
\end{align}
\end{subequations}
The unknown amplitudes $\xi_1$ and $\xi_2$ can be eliminated by solving \eqref{eqn:h1BC_lin_left} and \eqref{eqn:h1BC_lin_right}. The explicit expression for the boundary conditions then become
\begin{subequations}\label{eqn:h1BC_lin2}
\begin{align}
\hat{h}_1'(r_1) &= 0,\\
\hat{h}'_1(r_1+r_{w})&=\hat{h}_1(r_1+r_{w})\frac{h''_0(r_1+r_{w})}{h'_0(r_1+r_{w})}. 
\end{align}
\end{subequations}
For a given $n$, to solve \eqref{eqn:L1h1}, we map the $r$ space ($r_1 \leq r \leq r_1+r_{w}$) onto the $\zeta$ space ($-1 \leq \zeta \leq 1$) by
\begin{equation}
r=r_1+\frac{r_{w}}{2}(\zeta+1),\\
\label{eqn:r2zeta}
\end{equation}
and correspondingly $\hat{h}_1(r)$ is replaced by $g(\zeta)$. This gives the eigenvalue problem \eqref{eqn:L1h1} in terms of $\zeta$,
\begin{equation}
\mathcal{L}(g)=\mathrm{i}\lambda g,\\
\label{eqn:L1g}
\end{equation}
and $g(\zeta)$ can be discretized as 
\begin{equation}
g(\zeta)=\sum_{i=1}^{N}\beta_i\varphi_i,
\label{eqn:discreteg}
\end{equation}
where $\varphi_i$ are an orthogonal basis and $\beta_i$  are the spectral coefficients. To make $\varphi_i$ satisfy the boundary conditions at $\zeta=\pm1$, a linear combination of Chebyshev functions $T_i$ is adopted to form $\varphi_i$,
\begin{equation}
\varphi_i = T_{3i-3} + a_i T_{3i-2} + b_i T_{3i-1},
\label{eqn:varphi}
\end{equation}
where the coefficient $a_i$ and $b_i$ can be determined by substituting \eqref{eqn:varphi} into boundary conditions \eqref{eqn:h1BC_lin2}. The Gauss-Lobatto grid is adopted to discretize the radial space
\begin{equation}
\zeta_i =\cos\left(\frac{\pi i}{N-1}\right), i=1, 2, ..., N-2
\label{eqn:gridpoint}
\end{equation}
Finally, \eqref{eqn:L1h1} is transformed into a generalized matrix eigenvalue problem,
\begin{equation} \label{eqn:eigenEqn}
U\beta=\lambda V \beta, \ \ \  
\end{equation}
where $U_{i,j} = \mathcal{L}(\zeta_i, \varphi_j) $ and $V_{i, j}=\varphi_j(\zeta_i)$.

\subsection{Perturbation dynamics} 
\begin{figure}
  \centerline{\includegraphics[width=13cm]{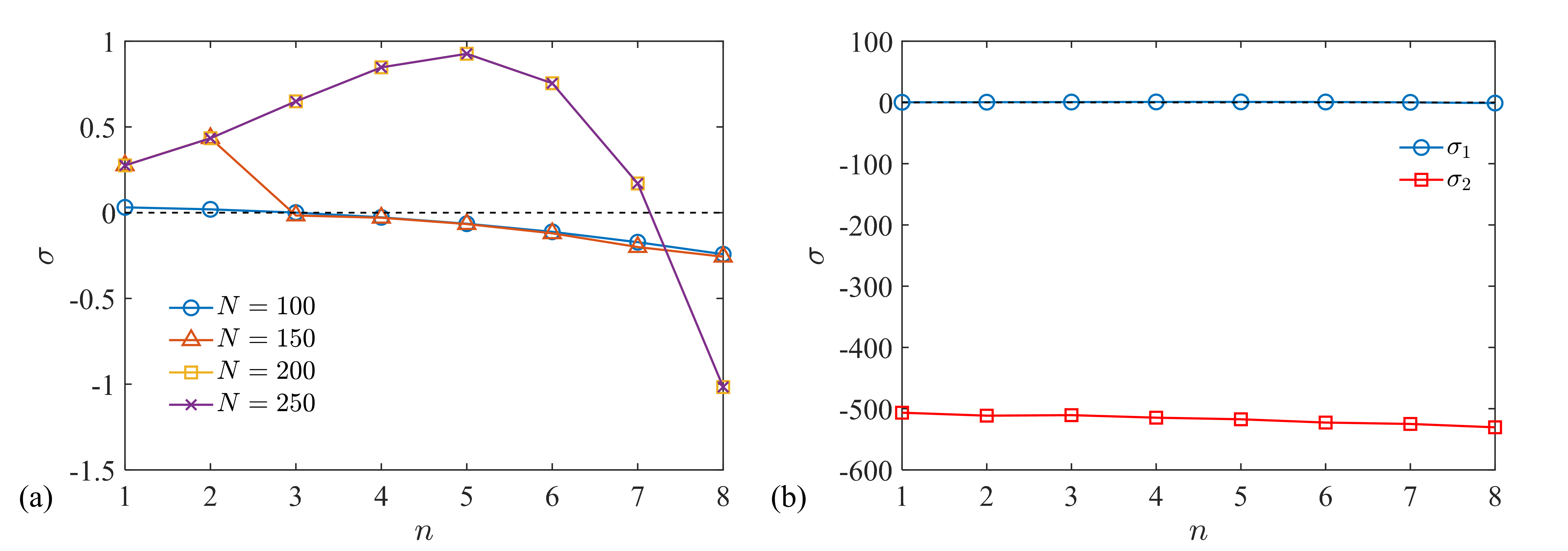}}
\caption{(a) Resolution sensitivity of the largest $\sigma$ on the grid number $N$ for $r_{w0}=0.26$, $\theta_1=90^{\circ}$, and $\theta_2=30^{\circ}$. (b) The corresponding largest and second largest growth rate, i.e., $\sigma_1$ and $\sigma_2$ vs. $n$ with $N=200$.}
\label{fig:EigenvalueTest}
\end{figure}

By solving the eigenvalue problem, the dependence of $\lambda$ on $n$ is obtained. Since the solved phase speed $\omega$ of the dominating perturbation modes equals zero, indicating that they are constant periodic structures that can grow or decay, we only discuss the growth rate $\sigma$ for the following. A resolution sensitivity test determined that $N=200$ is required to guarantee a convergent solution, as shown in figure \hyperref[fig:EigenvalueTest]{\ref{fig:EigenvalueTest}(a)}. Additionally, figure \hyperref[fig:EigenvalueTest]{\ref{fig:EigenvalueTest}(b)} shows that the second largest $\sigma$ is negative and much smaller than the first one. Therefore, only the largest $\sigma$ determines the film stability and is the one we focus on. For what follows, $\sigma$ represents the largest growth rate unless otherwise specified. 
The growth rate curve, $\sigma$ vs. $n$, quantifies the stability of the film-cylinder system to various perturbations. Positive $\sigma$ indicates exponential growth of a perturbation while a negative growth rate means that the film is stable with respect to the imposed perturbation. Zero ($0$, $n_{\mathrm{zero}}$) and peak ($\sigma_{\max}$, $n_{\max}$) growth rates are used to characterize the film stability. Specifically, the zero point, $n_{\mathrm{zero}}$, corresponds to the maximum possible number of satellite droplets for a given baseflow. Should $\sigma$ become negative and $n>n_{\mathrm{zero}}$, the corresponding perturbation would become suppressed. As such, $n_{\mathrm{zero}}$ defines the boundary between the stable and unstable regime. The peak growth rate, $\sigma_{\max}$, corresponds to $n_{\max}$ and indicates that the perturbation would grow the fastest and have the largest likelihood of dominating the instability process. Therefore, $n_{\max}$ is approximately the expected number of satellite droplets when the film is stimulated by a random perturbation containing a wide range of wavenumbers.

\begin{figure}
  \centerline{\includegraphics[width=13cm]{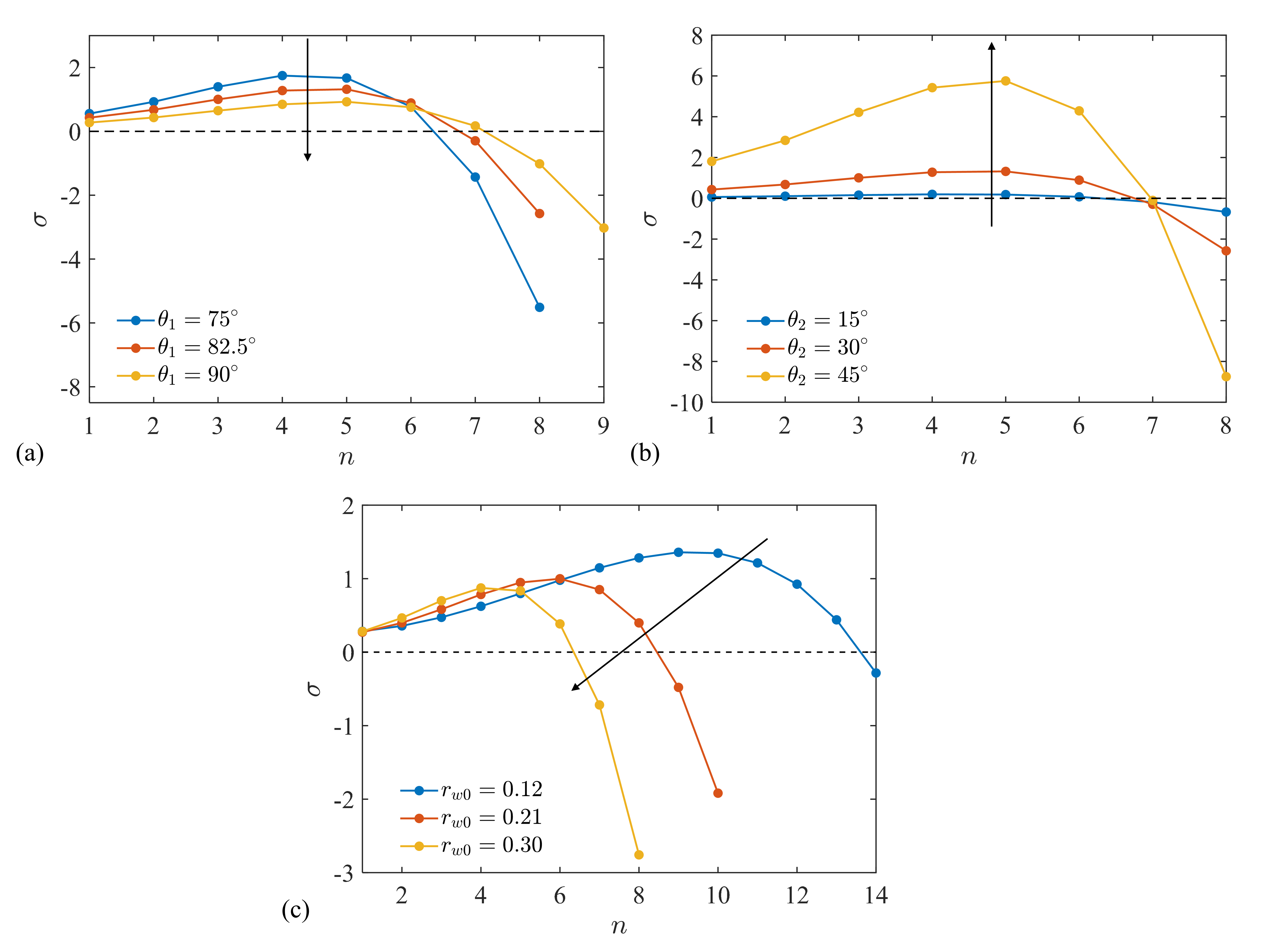}}
  \caption{Growth rate curves for (a) $r_{w0}=0.26$, $\theta_2 = 30^{\circ}$, and $\theta_1 = 75^{\circ}, 82.5^{\circ}, 90^{\circ}$;(b) $r_{w0}=0.26$, $\theta_1 = 82.5^{\circ}$, and $\theta_2 = 15^{\circ}, 30^{\circ}, 45^{\circ}$; and (c) $\theta_1 = 90^{\circ}$, $\theta_2 = 30^{\circ}$, and $r_{w0}=0.12, 0.21, 0.30$. The arrows show the direction of increasing $\theta_1$ in (a), $\theta_2$ in (b) and $r_{w0}$ in (c).}
\label{fig:dispersionRelation}
\end{figure}
Figure \ref{fig:dispersionRelation} shows the growth rate curves for the corner conditions shown in figure \ref{fig:h0}. 
Regarding the effect of the film size, as shown in figure \hyperref[fig:dispersionRelation]{\ref*{fig:dispersionRelation}(c)}, with increasing $r_{w0}$, both $n_{\max}$ and $n_{\mathrm{zero}}$ decline, suggesting that a thick corner film is more susceptible to low-wavenumber perturbations. Increasing $r_{w0}$ would weaken the relative difference between the curvatures at the crest, $1/R^c_2$, and the neck, $1/R^n_2$, which drives the film to break up, since the corner film becomes flatter as the film size becomes larger (see figure \hyperref[fig:h0]{\ref*{fig:h0}a}). However, the curvature difference of $1/R^c_1$ and $1/R^n_1$, which suppresses the instability, is comparatively less influenced by $r_{w0}$. Thus, the thicker the film becomes, the more stable it would be.

\begin{figure}
  \centerline{\includegraphics[width=13cm]{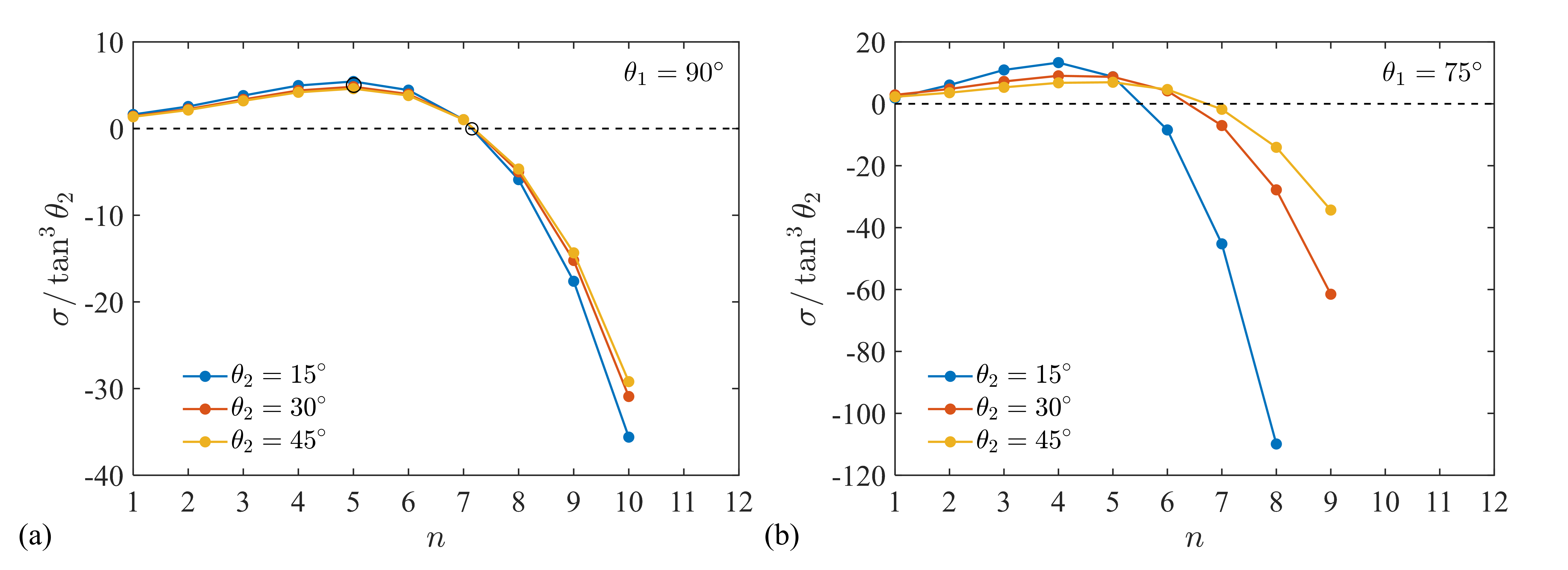}}
  \caption{Scaled growth rate  $\sigma /\tan^3\theta_2$ for the cases with $r_{w0}=0.26$, (a) $\theta_1 = 90^{\circ}$ and (b) $\theta_1 = 75^{\circ}$. The peak and zero points are marked by black circles in (a)}
\label{fig:scaling}
\end{figure}
Regarding the wettability effects, as $\theta_1$ increases from $75^{\circ}$ to $90^{\circ}$ in figure \hyperref[fig:dispersionRelation]{\ref*{fig:dispersionRelation}(a)}, the growth rate curve tends to be flatter, suggesting that $n_{\max}$ and $n_{\mathrm{zero}}$ increase while $\sigma_{\max}$ decreases. Contrary to the effect of $\theta_1$, $\sigma_{\max}$ increases with $\theta_2$, as shown in figure \hyperref[fig:dispersionRelation]{\ref*{fig:dispersionRelation}(b)}. \cite{gonzalez2013stability} suggested an approximate scaling $\sigma \propto \tan^3 \theta_2$ for a liquid ring on a solid substrate, which applies to cases where $\theta_1 = 90^{\circ}$ as shown in figure \hyperref[fig:scaling]{\ref*{fig:scaling}(a)}. However, figure \hyperref[fig:scaling]{\ref*{fig:scaling}(b)} shows that this scaling does not hold once $\theta_1$ takes other values, e.g. $75^{\circ}$. Since the eigenvalue problem is governed by the coefficients $\hat{c}_i$ in \eqref{eqn:L1} and $\hat{c}_i \sim h_0^3$, the scaling would require that $h_0 \sim \tan\theta_2$. According to \eqref{eqn:steadysol_paras}, this requirement will only be satisfied when $\theta_1 \to 90^{\circ}$ and therefore $\cot \theta_1 \to 0$. 

\subsection{Characterization of film stability}
A complete picture for characterizing the film stability in the parameter space $\theta_1\in\left[75^{\circ}, \ 90^{\circ}\right]$, $\theta_2\in\left[15^{\circ}, \ 45^{\circ}\right]$, and $r_{w0}\in\left[0.12,\ 0.30\right]$ can now be provided to answer the questions posed at the beginning. Namely, under what conditions does the corner film become unstable and how many satellite droplets are formed once the instability occurs.

\begin{figure}
  \centerline{\includegraphics[width=13cm]{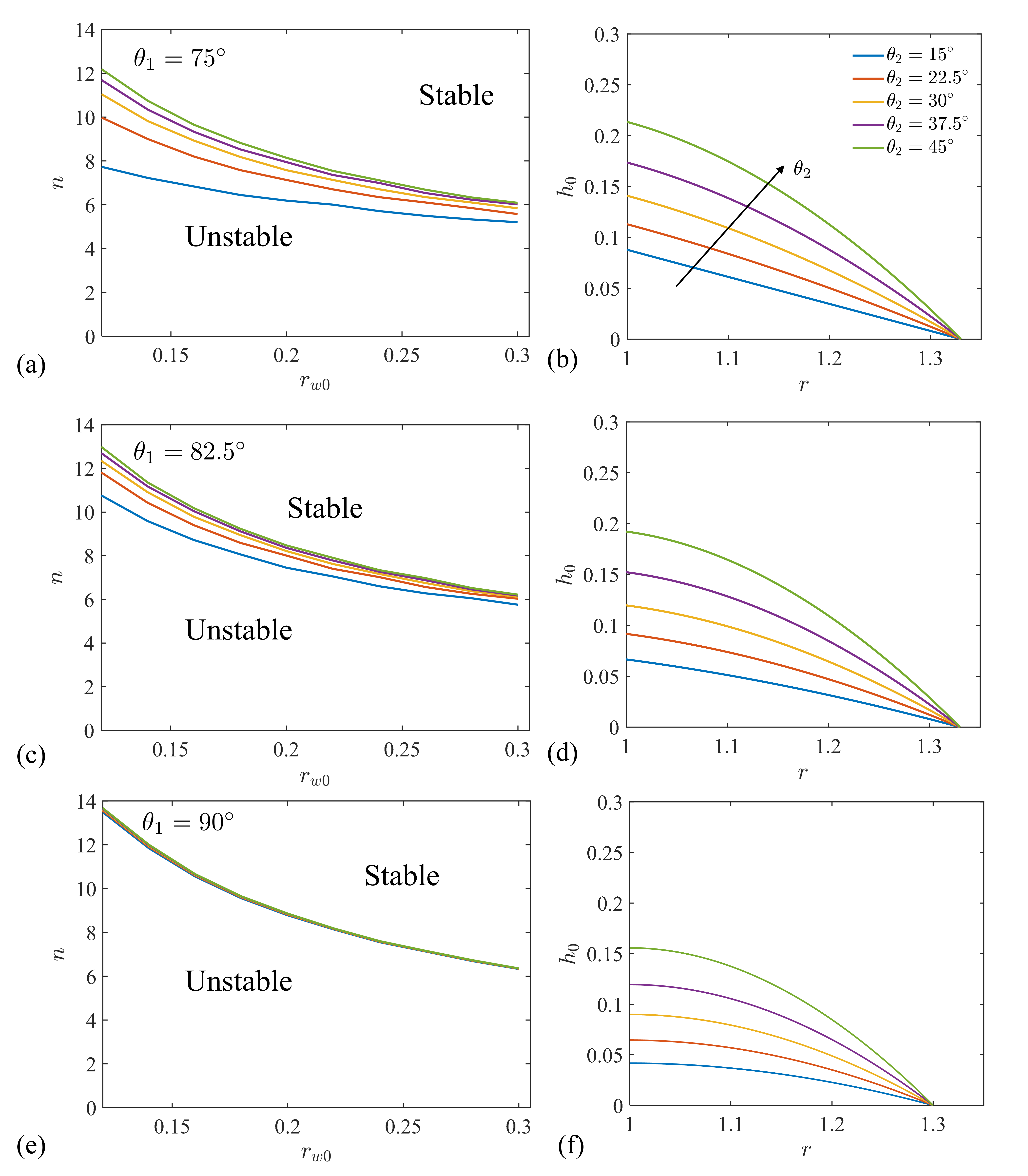}}
  \caption{Stability curves for $\theta_1=$ $75^{\circ}$(a-b), $82.5^{\circ}$(c-d) and $90^{\circ}$(e-f). Left column:  $n_{\mathrm{zero}}$-$r_{w0}$ curves with various $\theta_2 \in [15^{\circ}, 22.5^{\circ}, 30^{\circ},37.5^{\circ},45^{\circ}]$. Right column: equilibrium profiles with $r_{w0} = 0.30$ and various $\theta_2$.}
\label{fig:pd_n0}
\end{figure}

We first investigate the marginal stability corresponding to the critical condition under which the film would become unstable. This can be characterized by $n_{\mathrm{zero}}$ and $\sigma_{\mathrm{max}}$. The solid lines in figures \hyperref[fig:pd_n0]{\ref*{fig:pd_n0}(a)}, \hyperref[fig:pd_n0]{\ref*{fig:pd_n0}(c)}, and \hyperref[fig:pd_n0]{\ref*{fig:pd_n0}(e)} depict $n_{\mathrm{zero}}$ vs. $r_{w0}$ and distinguish the regions of stability (upper) and instability (lower). The stable region is modified by the wall wettability. As $\theta_1$ and $\theta_2$ decrease, the stable region becomes enlarged. In particular, for $\theta_1=90^{\circ}$, the condition corresponding to $n_{\mathrm{zero}}$ becomes independent of $\theta_2$ and the curves for various $\theta_2$ collapse as one in figure \hyperref[fig:pd_n0]{\ref*{fig:pd_n0}(c)} owing to the scaling law previously mentioned in connection with figure \hyperref[fig:scaling]{\ref*{fig:scaling}(a)}. With increasing $r_{w0}$, $n_{\mathrm{zero}}$ decreases and eventually approaches a value of $n=6$ or $n=7$. This seems to suggest that the corner film tends to be more stable when it becomes thicker, but full stability cannot be achieved. However, due to the assumption of the long-wave theory, i.e. $r_{w0} \ll r_1$, the LSA may not lead to accurate predictions if the corner film is too thick. 

\begin{figure}
  \centerline{\includegraphics[width=13cm]{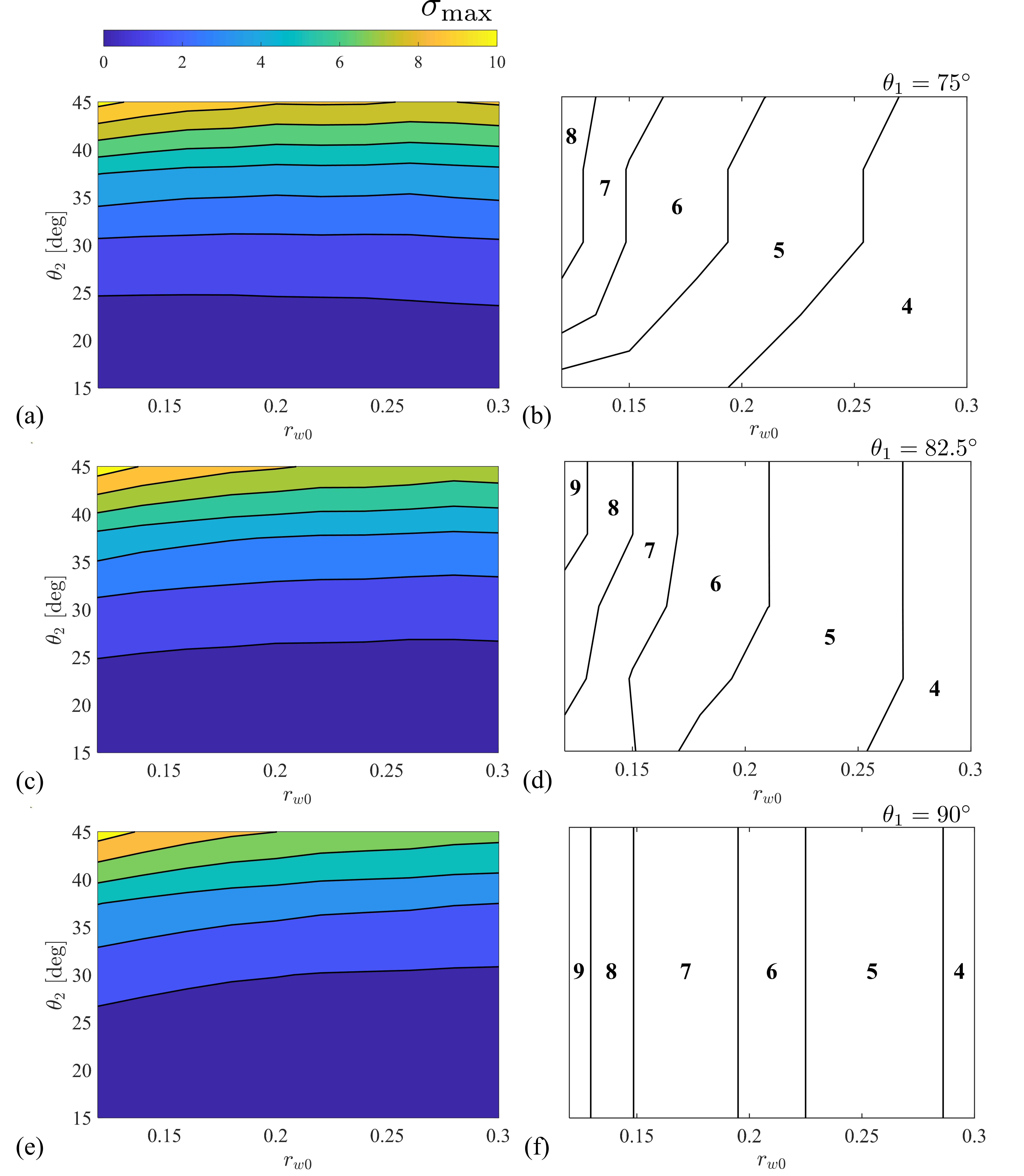}}
  \caption{Contours of $\sigma_{\mathrm{max}}$ (a, c, e) and $n_{\mathrm{max}}$ (b, d, f) for $\theta_1=$ $75^{\circ}$(a-b), $82.5^{\circ}$(c-d), and $90^{\circ}$(e-f) in the parameter space of $\theta_2$ and $r_{w0}$. The values of contour lines in (a, c, e) are $[1, 2, 3, 4, 5, 6, 7, 8]$.}
\label{fig:pd_nmax}
\end{figure}

Focus is now placed on predictions of the expected number of satellite droplets, which correspond to the perturbation mode with the maximum growth rate. Figure \ref{fig:pd_nmax} shows the contours of $\sigma_{\mathrm{max}}$ and $n_{\mathrm{max}}$, with the latter indicating the expected number of satellite droplets. Note that $\sigma_{\mathrm{max}}$ is mainly controlled by wall wettability and only slightly affected by the film size. Figures \hyperref[fig:pd_nmax]{\ref*{fig:pd_nmax}(b)}, \hyperref[fig:pd_nmax]{\ref*{fig:pd_nmax}(d)}, and \hyperref[fig:pd_nmax]{\ref*{fig:pd_nmax}(f)} show that the expected number of satellite droplets generally decreases with $r_{w0}$ while ranging from 4 to 9.

Overall, the film stability can be characterized by the growth rate curve obtained from the LSA. Specifically, the post-instability pattern, including the maximum and most probable number of satellite droplets following film breakup, can be predicted by $n_{\mathrm{zero}}$ and $n_{\mathrm{max}}$, respectively. The film size $r_{w0}$ plays the most important role in determining the marginal stability and the post-instability pattern of a corner film. The thinner the film is, the more susceptible it becomes to perturbations of a larger wavenumber. The wall wettability including $\theta_1$ and $\theta_2$ mainly influence $\sigma_{\mathrm{max}}$ while having a secondary influence on $n_{\mathrm{zero}}$ and $n_{\mathrm{max}}$. In the following sections, we will compare the LSA prediction against the numerical results obtained using the disjoining pressure model (DPM) and volume-of-fluid (VOF) simulations to assess to what extent can the LSA predict the post-instability development, especially when nonlinear effects become involved in the breakup stage. 

\section{Disjoining pressure model (DPM)} \label{sec:DPM}
\begin{figure}
  \centerline{\includegraphics[width=10cm]{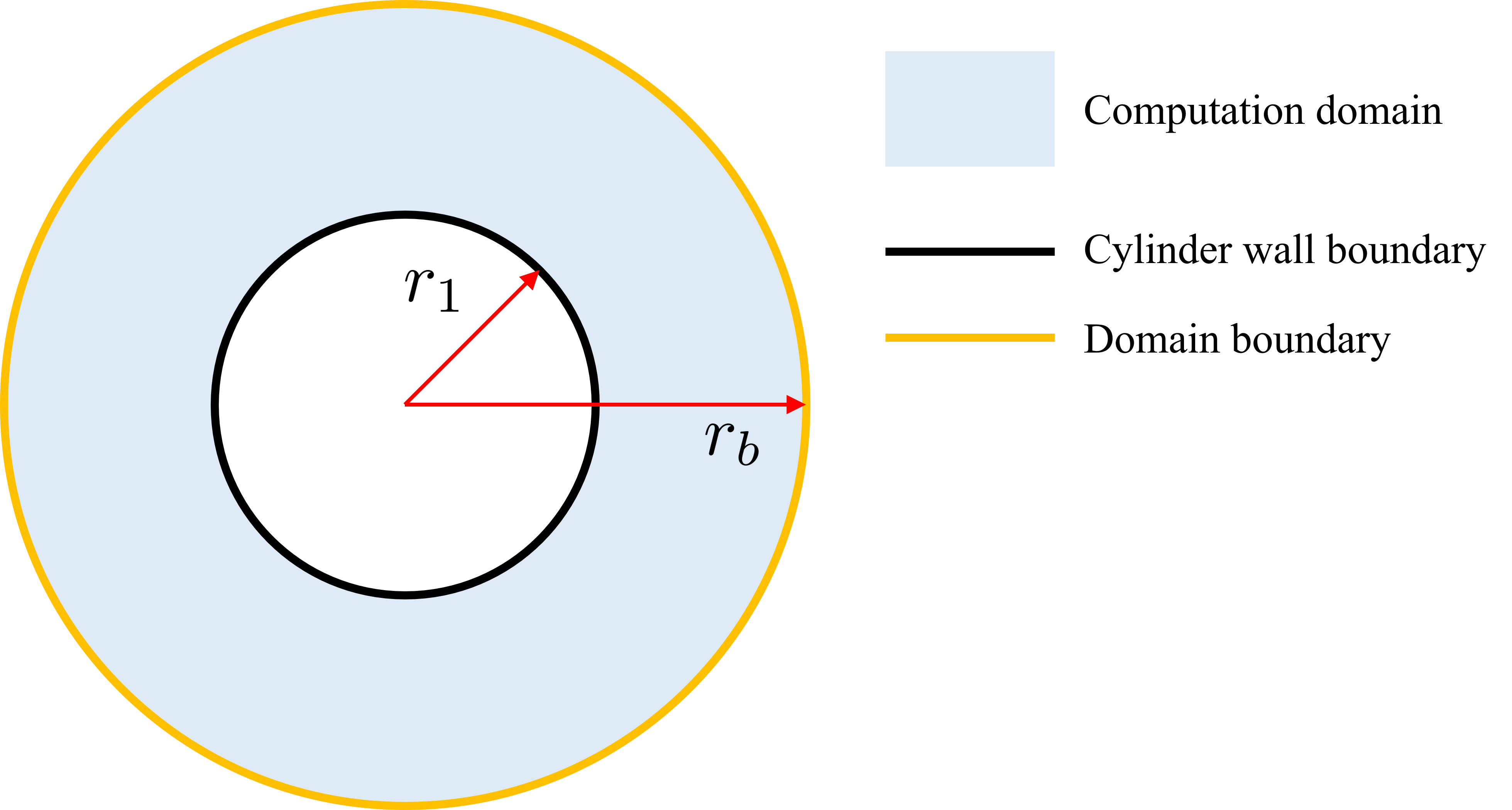}}
  \caption{The schematic of the numerical model for the DPM, including a computation domain lying between the cylinder wall and the domain boundary. Here, $r_1$ is the cylinder radius and $r_b$ defines the domain size.}
\label{fig:DPM_domain}
\end{figure}
Rather than using the Navier-slip boundary condition to address stress singularity at the contact line and directly applying contact angles, as was done in section \ref{LSA}, a precursor film is assumed and the wettability condition on the walls can be represented via a disjoining pressure $\Pi(h)$ \citep{kondic2020liquid, diez2007breakup, savva2011dynamics},
\begin{equation}\label{eqn:DPMeq_0}
\frac{\partial h}{\partial t}+\nabla \cdot\left[h^{3} \nabla \nabla^{2} h+h^{3}\nabla\Pi(h)\right]=0.
\end{equation}
The most commonly used form of $\Pi(h)$ is adopted here \citep{mitlin1994nonlinear, schwartz1998hysteretic},
\begin{equation}\label{eqn:disjoiningModel}
\Pi(h) = Kf(h)=K\left[ \left( \frac{h_*}{h} \right)^n- \left(\frac{h_*}{h} \right)^m\right],
\end{equation}
where $f(h)$ represents liquid-solid repulsion and attraction with exponents $n>m>1$. Throughout the literature, the dynamic effects of different exponent pairs $(n, m)$ including $(9, 3)$, $(4, 3)$, and $(3,2)$ have been discussed \citep{craster2009dynamics,kondic2020liquid}. In this work, we adopt $(3,2)$ \citep{schwartz1998simulation}.
This liquid-solid interaction leads to a thickness $h_*$ at which the repulsive and attractive forces are balanced. $h_*$ is related to a precursor film thickness \citep{savva2011dynamics}, which under realistic conditions is of nanoscale thickness. Hence, $h_*$ should be extremely small. Computationally, the grid spacing must also be close to $h_*$ to guarantee numerical convergence. To maintain a reasonable level of computational cost, we adopt $h_*=10^{-3}$. The constant $K=2(1-\cos \theta_*)/h_*$, where $\theta_*$ is related to the contact angle. Due to interface relaxation, the contact angle measured from the equilibrium state would be smaller than $\theta_*$. Finally, the governing equation \eqref{eqn:heqn} becomes
\begin{equation}\label{eqn:DPMeq}
\frac{\partial h}{\partial t}+\nabla \cdot\left[h^{3} \nabla \nabla^{2} h\right]+
K\nabla \cdot \left(h^3 f'(h)\nabla h\right)=0.
\end{equation}
Equation \eqref{eqn:DPMeq} is solved for the domain shown in figure \ref{fig:DPM_domain} with the following boundary condition on the cylinder wall
\begin{equation}\label{eqn:DPMeqBC1}
\nabla h \cdot \boldsymbol{n} = -\cot \theta_1,
\end{equation}
where $\boldsymbol{n}$ is the unit normal vector, and on the domain boundary
\begin{equation}\label{eqn:DPMeqBC2}
h  = h_*.
\end{equation}
To prevent the outer boundary from affecting the film evolution, the domain size, $r_b$, is set two times larger than $r_1$.

\begin{figure}
\centerline{\includegraphics[width=9.9cm]{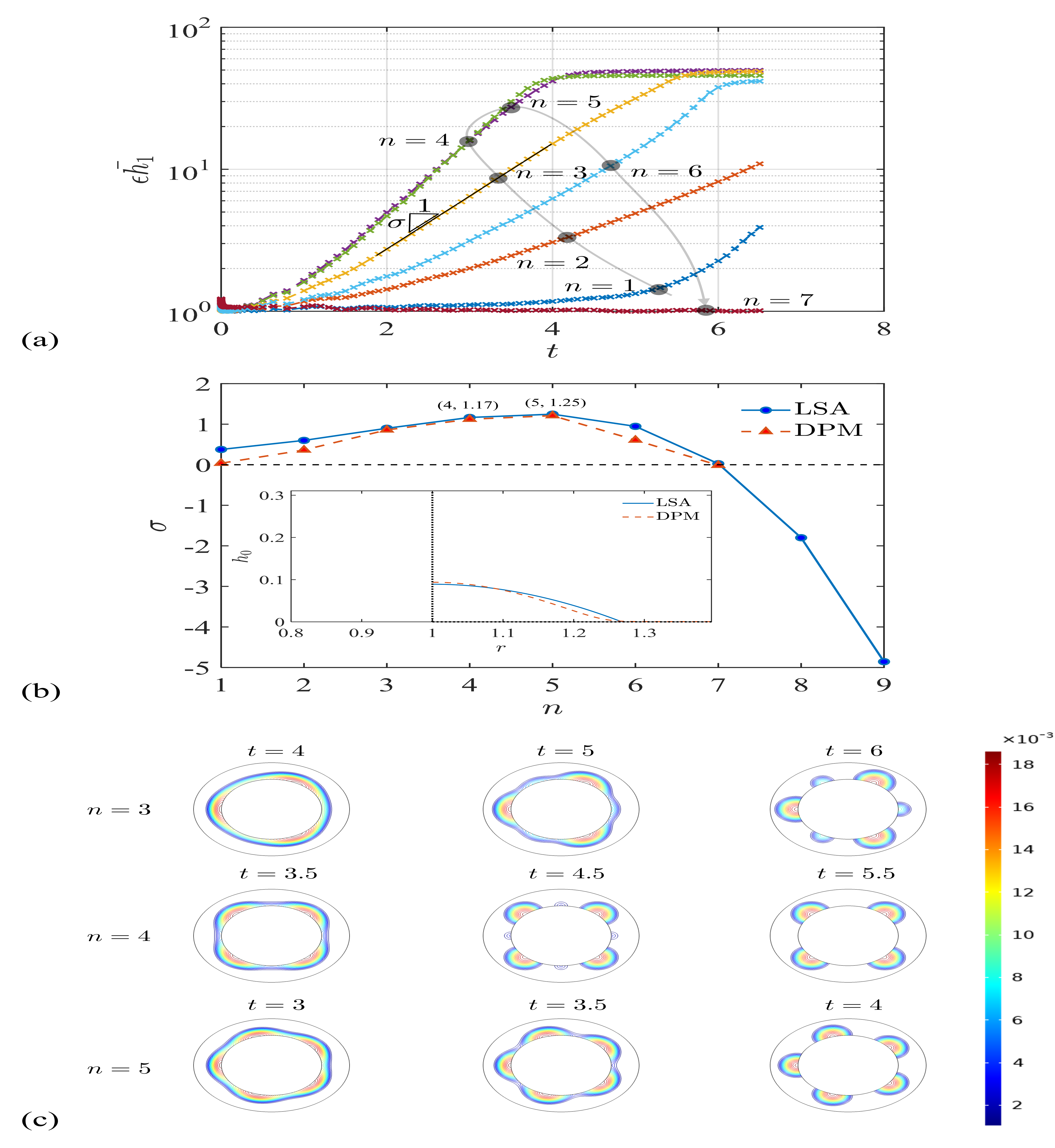}}
  \caption{(a) The perturbation growth for the case  with $r_{w0}=0.27$, $\theta_1 = 90^{\circ}$ and $\theta_* = 60^{\circ}$. (b) The comparison of growth rate between DPM and LSA. The inset in (b) shows the steady profile used in the DPM and correspondingly the equilibrium solution with $\theta_2 = 33^{\circ}$ for the LSA. (c) The evolution of film morphology from the instability occurrence to film breakup.}
\label{fig:Disjoining_h1}
\end{figure}

First, an equilibrium profile needs to be obtained. Specifically, a solution obtained from \eqref{eqn:steadysol} is elevated by $h_*$, thereby guaranteeing that $h\geq h_*$. This modified solution is taken as an initial profile. It evolves by solving \eqref{eqn:DPMeq} in an axisymmetric domain until a steady state is achieved. The final steady solution may be different from the initial profile due to the contact angle relaxation. To match the LSA and DPM, it is necessary to re-evaluate the corner condition used in the LSA to conform to the steady solution from the DPM. We keep the liquid volume $V$, $r_{w0}$ and $\theta_1$ fixed, then adjust $\theta_2$ to fit the steady profile of the DPM. One case is shown in the inset of figure \hyperref[fig:Disjoining_h1]{\ref*{fig:Disjoining_h1}(b)} where initially $r_{w0}=0.27$, $\theta_1 = 90^{\circ}$ and $\theta_* = 60^{\circ}$, but the final fitting yields $\theta_2 = 33^{\circ}$.

Once the equilibrium profile is obtained, we add a perturbation $\epsilon h_1(\phi)|_{t=0}$ and solve Eq. \eqref{eqn:DPMeq} for
\begin{equation}
h(r, \phi)|_{t=0} =h_0(r)+\epsilon h_1(\phi)|_{t=0}.
\end{equation}
Finally, we obtain the time evolution of $h$, from which the growth rate $\sigma$ can be estimated as,
\begin{equation} \label{eqn: eps_h1_}
 \bar{\epsilon h_1}(t) = \sqrt{\int_{\Omega} \left[h(r, \phi, t)-h_0(r)\right]^2 d\Omega }.
\end{equation}
Here, $\Omega$ is the computation domain as shown in figure \ref{fig:DPM_domain}.

\subsection{Single-mode perturbations}

We start by investigating the dynamics of a corner film triggered by single-mode perturbations, i.e.
\begin{equation}
\epsilon h_1(\phi)|_{t=0} =A_n \cos (n\phi).
\end{equation}
The perturbation amplitude $A_n = 10^{-4}$ is small enough to guarantee that linearity dominates in the initial stage. The evolution of $\bar{\epsilon h_1}$ for $n=1,...,7$ are shown in figure \hyperref[fig:Disjoining_h1]{\ref*{fig:Disjoining_h1}(a)}. After a short oscillation period, the perturbation grows exponentially and thus the growth rate for each wavenumber can be obtained by linear fitting. As shown in figure \hyperref[fig:Disjoining_h1]{\ref*{fig:Disjoining_h1}(b)}, results of the LSA and DPM are in good agreement around $n_{\mathrm{max}}$ and demonstrate a consistent trend. We do not expect a perfect quantitative agreement since the contact line modeling on the bottom wall differs. The contact line is an asymptotic slope of the interface in the DPM, while a specific contact angle is set in the LSA.

The film breakup process exhibits linear behavior initially, with the number of emerging satellite droplets being consistent with the wavenumber $n$, as is shown in the first column of figure \hyperref[fig:Disjoining_h1]{\ref*{fig:Disjoining_h1}(c)}. However, non-linearity dominates the final part of the evolution and results in a more complex film morphology. Taking $n=3$ as an example, a slim and long film forms at each neck region which connects two neighboring droplets. At $t=5$, the connecting film breaks up at its two ends and then forms a secondary droplet. Eventually, besides three major droplets, three secondary droplets appear in-between the major ones. For $n=4$, the connecting films become relatively shorter, resulting in smaller secondary droplets. Thus, the Laplace pressure of the secondary droplet becomes much larger than the one of the neighboring major droplet, and this pressure difference drives the liquid to fast flow from the secondary droplets to the major ones. Eventually, secondary droplets only appear temporally and are absorbed by neighboring major ones. For $n=5$, the connecting film is too short to form a secondary droplet and only major satellite droplets appear during the breakup.

\subsection{Random perturbations}

\begin{figure}
  \centerline{\includegraphics[width=13cm]{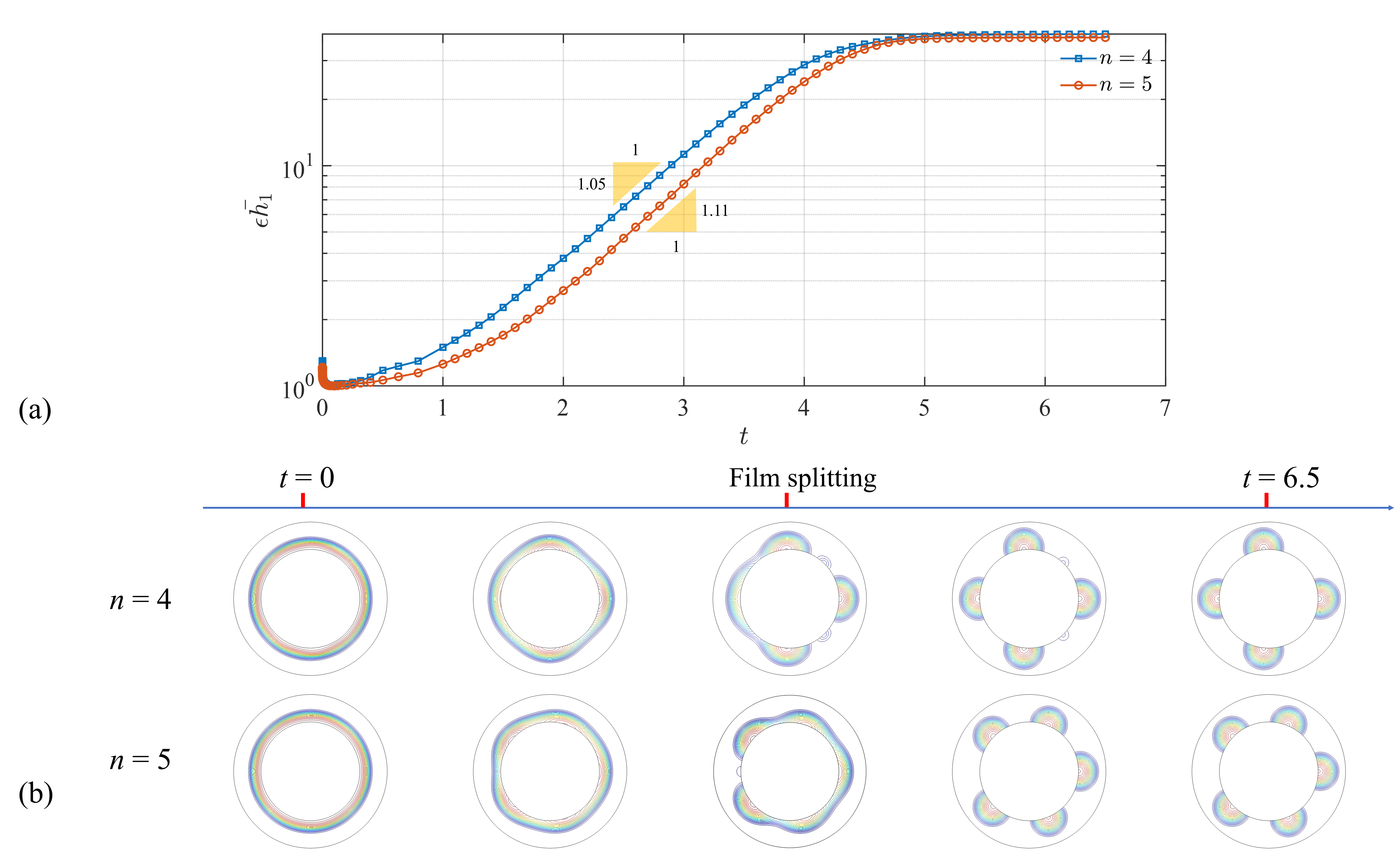}}
  \caption{(a) The perturbation growth for two typical random perturbations leading to the mode $n=4$ and $n=5$; (b) the snapshots of the evolution of the film morphology.}
\label{fig:mixedperturbation}
\end{figure}

Generally, perturbations are of a random nature covering a wide range of wavenumbers. We consider a superposition of $N$ individual single-mode perturbations,
\begin{equation}
\epsilon h_1(\phi)|_{t=0} =\sum_{n=0}^{N} A_n \cos (n\phi),
\end{equation}
where $A_n$ are random amplitudes within $[-A_{max}, \ A_{max}]$ with $A_{max}=10^{-4}$ and $N=100$. Ten samples of $\epsilon h_1(\phi)|_{t=0}$ are generated and superimposed on the baseflow shown in the inset of figure \hyperref[fig:Disjoining_h1]{\ref*{fig:Disjoining_h1}(b)}. According to the LSA, the perturbation mode with $n=5$ has the maximum growth rate and thus the film is expected to break up into five satellite droplets. However, for this case, the four-droplet and the five-droplet pattern are equally likely to occur since the difference between the first and second largest growth rate is small, i.e. $\sigma = 1.16$ for $n=4$ while $\sigma = 1.25$ for $n=5$. Perturbation energy obtained from the DPM is shown in figure \hyperref[fig:mixedperturbation]{\ref*{fig:mixedperturbation}(a)} for the two perturbations. Snapshots of the corner film evolving after being perturbed are presented in figure \hyperref[fig:mixedperturbation]{\ref*{fig:mixedperturbation}(b)}. Due to randomness, the film breakup at the neck regions is asynchronous, unlike single-mode perturbations, and therefore the final film morphology loses symmetry. Consequently, the connecting films have different lengths, which in turn causes the emerging secondary droplets to appear at random between two major droplets. For example, as shown in the third column of figure \hyperref[fig:mixedperturbation]{\ref*{fig:mixedperturbation}(b)} for $n=4$, only two secondary droplets temporarily appear as opposed to four appearing when the perturbation is single-mode. For $n=5$, one temporary secondary droplet forms while none appears for single-mode perturbation.

In summary, the DPM provides a direct insight into the dynamics of interfacial evolution. The LSA is quantitatively validated with the DPM results at the early stage. As the wrapping film approaches breakup, secondary droplets may appear between the major ones due to the formation of connecting films, especially when the perturbation wavenumber is small, like $n=3$ or $n=4$, and connecting films are long. Therefore, the LSA can provide a reliable prediction for the number of major droplets while the emergence of secondary droplets is beyond its capability.

\section{Volume of Fluid simulations} \label{sec:VOF}

\begin{figure}
  \centerline{\includegraphics[width=13cm]{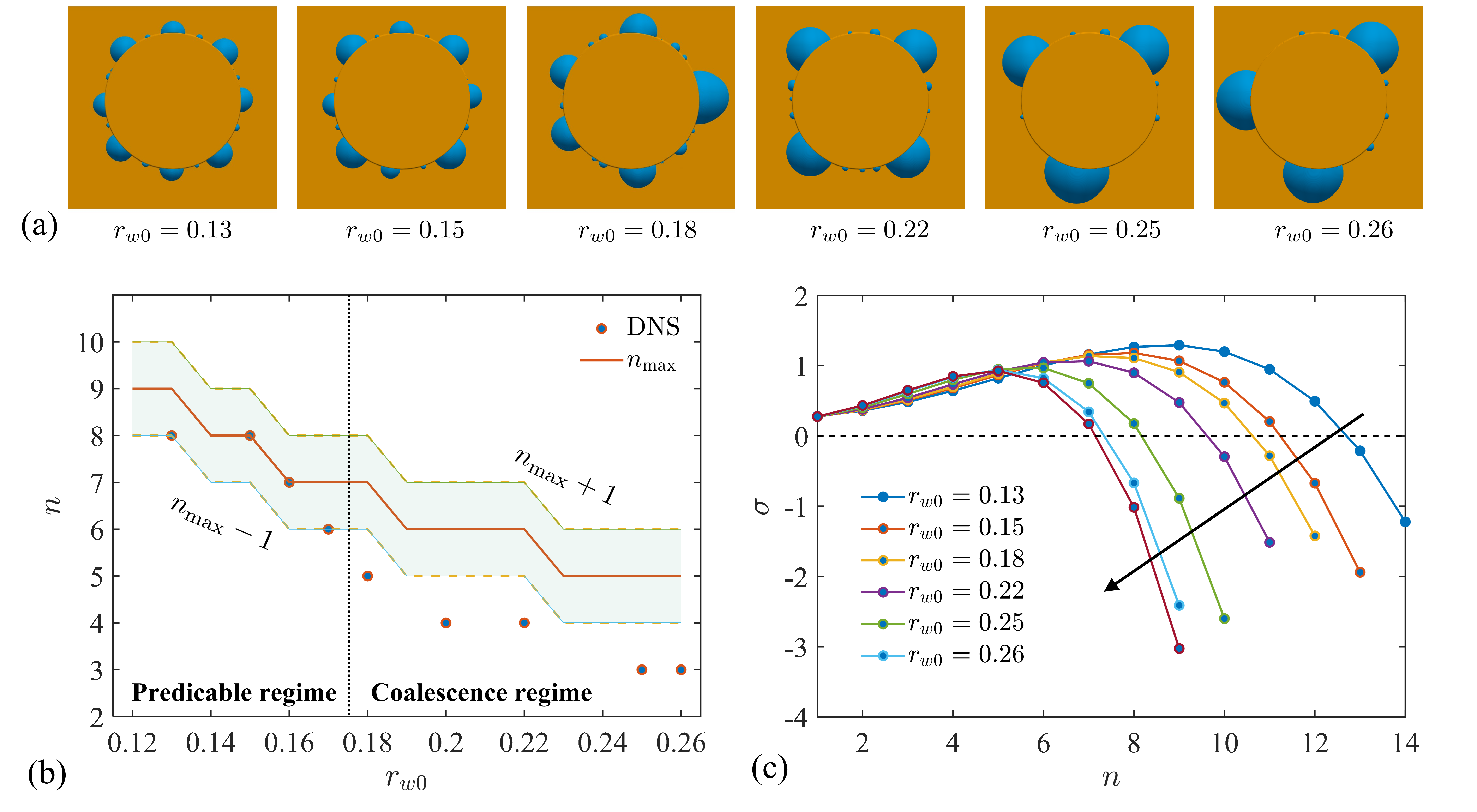}}
  \caption{(a) The final patterns of simulation cases with $\theta_1=90^{\circ}$, $\theta_2=30^{\circ}$ and varying film sizes $r_{w0}\in[0.13,0.26]$. (b) Number of satellite droplets obtained from VOF compared against the LSA predictions. (c) Growth rate curves of the corresponding cases.}
\label{fig:DNS_all}
\end{figure}

To further investigate the film breakup process and post-instability patterns, we conduct numerical simulations solving the incompressible Navier-Stokes equations and which eschew the assumptions employed in the LSA and DPM. The results of these simulations serve as an important reference against which the LSA and DPM results can be gauged, and will help assess how applicable the long-wave theory is to film stability issues. Specifically, we want to find the critical film size for which the lubrication approximation gives good predictions of the post-instability patterns of a wrapping film.

The finite-volume solver used to conduct the VOF simulations is the open-source code FluTAS \citep{crialesi2022flutas}. It employs the MTHINC algebraic volume-of-fluid method \citep{ii2012interface} for the numerical realization and advection of the interface between two immiscible fluids. To impose contact angles on solid geometries of arbitrary shape, the code is combined with the ghost-cell immersed-boundary method of \citet{shahmardi2021fully}, which allows us to use the extrapolation procedure proposed by \citet{RENARDY2001} for prescribing contact angles. The setup used with this code, shown in figure \hyperref[fig:f0]{\ref*{fig:f0}(a)}, consists of a domain with dimensions of $[L_x, L_y, L_z] = [1.0, 1.0, 0.3]$ where a cylinder is mounted on top of a flat wall. The domain boundaries are periodic along $x$ and $y$ with no-slip and impenetrability imposed at the boundaries along $z$. The uniform grid spacing is $[\Delta x, \Delta y, \Delta z] = [L_x/N_x, L_y/N_y, L_z/N_z]$ with $[N_x, N_y, N_z] = [500, 500, 150]$. In its initial state, the liquid film rests on the flat wall and wraps around the cylinder, with the shape of its profile defined using \eqref{eqn:steadysol}. The simulations were ran until a steady state was achieved and the liquid film underwent no further evolution. Note that unlike the artificial perturbation used in the DPM, the perturbation in the VOF simulations originates from numerical sources. Specifically, this would be the ``roughness'' of the cylinder wall, due to the fact that the immersed boundaries in the Cartesian coordinate system are not perfect representations of actual arcs. However, due to the fine grid resolution adopted, the amplitude of the induced perturbation is low enough to guarantee the linear-perturbation assumption. The grid dependence results are gathered in appendix \ref{app:grid}.

The post-instability patterns of simulations with film sizes of $r_{w0}\in[0.13, 0.26]$ and fixed $\theta_1=90^{\circ}$, $\theta_2=30^{\circ}$ are demonstrated in figure \hyperref[fig:DNS_all]{\ref*{fig:DNS_all}(a)}. In addition to film break-up, major satellite droplets form while sub-droplets sometimes randomly occur (additional details in appendix \ref{app:grid}). If we ignore the sub-droplets and focus on the major satellite droplets, the VOF simulations can be compared against the $n_{\mathrm{max}}$ predictions from the LSA. Since the growth rates of $n_{\mathrm{max}}-1$, $n_{\mathrm{max}}$ and $n_{\mathrm{max}}+1$ are close, as shown in figure \hyperref[fig:DNS_all]{\ref*{fig:DNS_all}(c)}, the corresponding post-instability patterns have a similar likelihood of appearing. Thus, the LSA prediction is regarded as reasonable when the corresponding VOF result lies in $[n_{\mathrm{max}}-1, n_{\mathrm{max}}+1]$. As shown in figure \hyperref[fig:DNS_all]{\ref*{fig:DNS_all}(b)}, there exists a critical film size $r^c_{w0}\approx0.175$ and highlighted by a dash line. If $r_{w0}<r^c_{w0}$, denoted the predictable regime, the LSA can provide a reasonable prediction which is in line with the VOF result. Otherwise, for $r_{w0}>r^c_{w0}$ which we denote the coalescence regime, the number of major satellite droplets is always smaller than the one predicted by the LSA. 

\begin{figure}
  \centerline{\includegraphics[width=13cm]{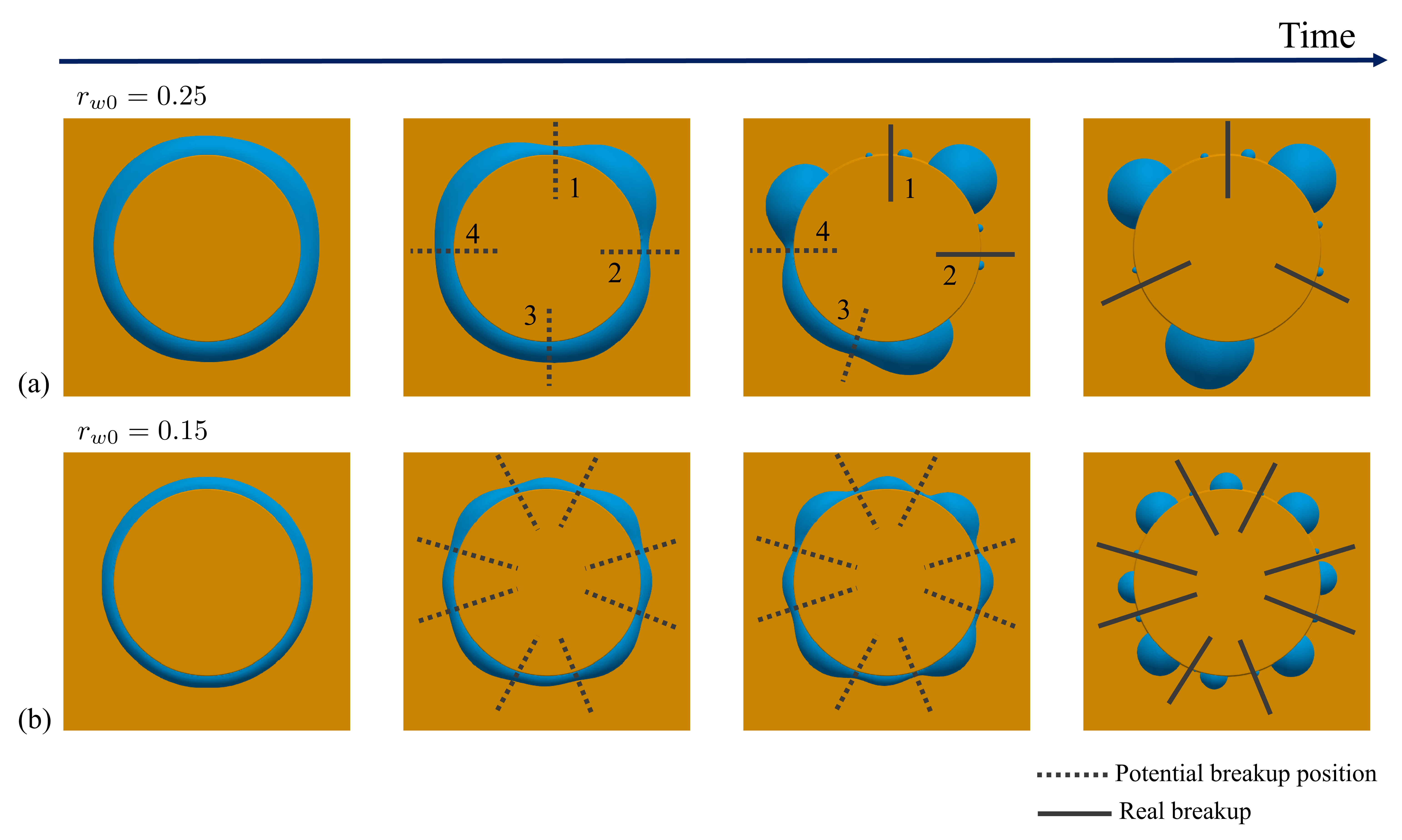}}
  \caption{The time evolution of two representative cases: (a) $r_{w0} = 0.25$ in the coalescence regime and (b) $r_{w0}=0.15$ in the predictable regime. For both cases, $\theta_1 = 90^{\circ}$ and $\theta_2 = 30^{\circ}$.}
\label{fig:DNS_two}
\end{figure}

Figures \hyperref[fig:DNS_two]{\ref*{fig:DNS_two}(a)} and \hyperref[fig:DNS_two]{\ref*{fig:DNS_two}(b)} demonstrate the time evolution of two representative cases belonging to the coalescence and predictable regime, respectively. For the case with $r_{w0}=0.25$ belonging to the coalescence regime, at the early stage, there appear four potential neck regions which are numbered in figure \hyperref[fig:DNS_two]{\ref*{fig:DNS_two}(a)}, indicating that mode $n=4$ dominates the early stages of the film evolution. This agrees with the LSA prediction, as shown in figure \hyperref[fig:DNS_all]{\ref*{fig:DNS_all}(b)}. However, the curvatures of these four necks are not uniform, resulting in different developing paces. Neck regions 1 and 2 which have a much larger curvature ($1/R^n_2$) become significantly squeezed due to the rather stronger capillary pressure and break up first. This is followed by neck region 4 breaking up, while the curvature of neck region 3 diminishes and gradually disappears, leading to the coalescence of the two neighboring crests on either side of this neck region. The asynchronous breakup of necks can be clearly observed in the coalescence regime. This asynchronous breakup induced by the different neck curvatures is also observed in the DPM results (figure \ref{fig:mixedperturbation}). Nevertheless, as the film becomes thinner, they fall into the predictable regime, and the neck curvatures tend to be similar which results in shorter neck-breaking times. As a result, crest coalescence is avoided and eight satellite droplets appear at the same time as shown in figure \hyperref[fig:DNS_two]{\ref*{fig:DNS_two}(b)} for the case of $r_{w0}=0.15$.

To conclude, the LSA can provide reasonable predictions that agree with the results from VOF simulations when the film is thin enough, i.e $r_{w0}<0.175$. However, for thick films, the final number of satellite droplets may not agree with the predictions, since crest coalescence becomes involved during the breakup of the neck regions, and its behaviour is not accounted for within the linear analysis framework. Most probably, the final number of satellite droplets should be smaller than the LSA prediction due to this coalescence mechanism. Therefore, $n_{\mathrm{max}}$ can be regarded as the upper limit of the number of satellite droplets. 

\section{Conclusions}

In this work, the stability of a corner film wrapping a cylinder is comprehensively studied. Theoretically, we develop a simplified model for describing the evolution of a corner film within the long-wave theory framework. Based on this model, LSA is performed to characterize the film stability and predict post-instability patterns. We find that the stability of the wrapping film is mainly controlled by the film size. In particular, the thicker the film is, the more sensitive it becomes to perturbations of a smaller wavenumber. As a consequence, fewer satellite droplets appear after film breakup. The wall wettability impacts the value of the growth rate but slightly influences marginal stability and post-instability patterns. Particularly when $\theta_1=90^{\circ}$, the film stability becomes independent of $\theta_2$ and only depends on $r_{w0}$. 

We introduced a disjoining pressure in the model so that the evolution of film morphology can be directly tracked numerically. The solution of the DPM not only served as a reference against which to validate the LSA, but to also help gain more understanding of the instability process. In the early stages of film evolution, the development follows the prediction of the LSA. Also, when the perturbation wavenumber becomes large, the wrapping film's break up is as predicted by the LSA. Otherwise, connecting films are formed at each neck region afterwards. These films ultimately break up at their ends, resulting in stable or temporary secondary droplets, the emergence of which depends on the length of the connecting film. Furthermore, numerical simulations using VOF are conducted. We find that crest coalescence may become involved during the breakup stage if the film is too thick ($r_{w0}>0.175$). This is due to the capillary pressure at the neck regions differing from one another. Due to this coalescence mechanism, the LSA prediction only provides an upper limit for the final number of satellite droplets.

Although this work is to a certain extent motivated by our previous work \citep{suo2022mobility} regarding liquid-stain removal in porous media, the conclusion of this work can also be leveraged for other engineering purposes, such as coating complex surfaces \citep{sabate2019antifouling} and self-assembly of liquid phases \citep{wu2014directed}. Furthermore, we expect that the knowledge of the film instability will enable us to realize passive control of the interfacial morphology through purposeful design of the geometry or topology. 

\section*{Acknowledgments}
This research was undertaken with the financial support of the Swedish Foundation for Strategic Research (SSF), provided through grant SSF-FFL15-001. Access to the supercomputing resources of the PDC Center for High Performance Computing and National Supercomputer Centre (NSC) used for this work were provided by the National Academic Infrastructure for Supercomputing in Sweden (NAISS).

\section*{Declaration of Interests}
The authors report no conflicts of interest.

\appendix
\counterwithin{figure}{section}
\counterwithin{table}{section}

\section{Grid dependence study} \label{app:grid}

\noindent
\begin{minipage}{\textwidth}
\makeatletter
\def\@captype{figure}
\makeatother
 \begin{center}
  \includegraphics[width=13cm]{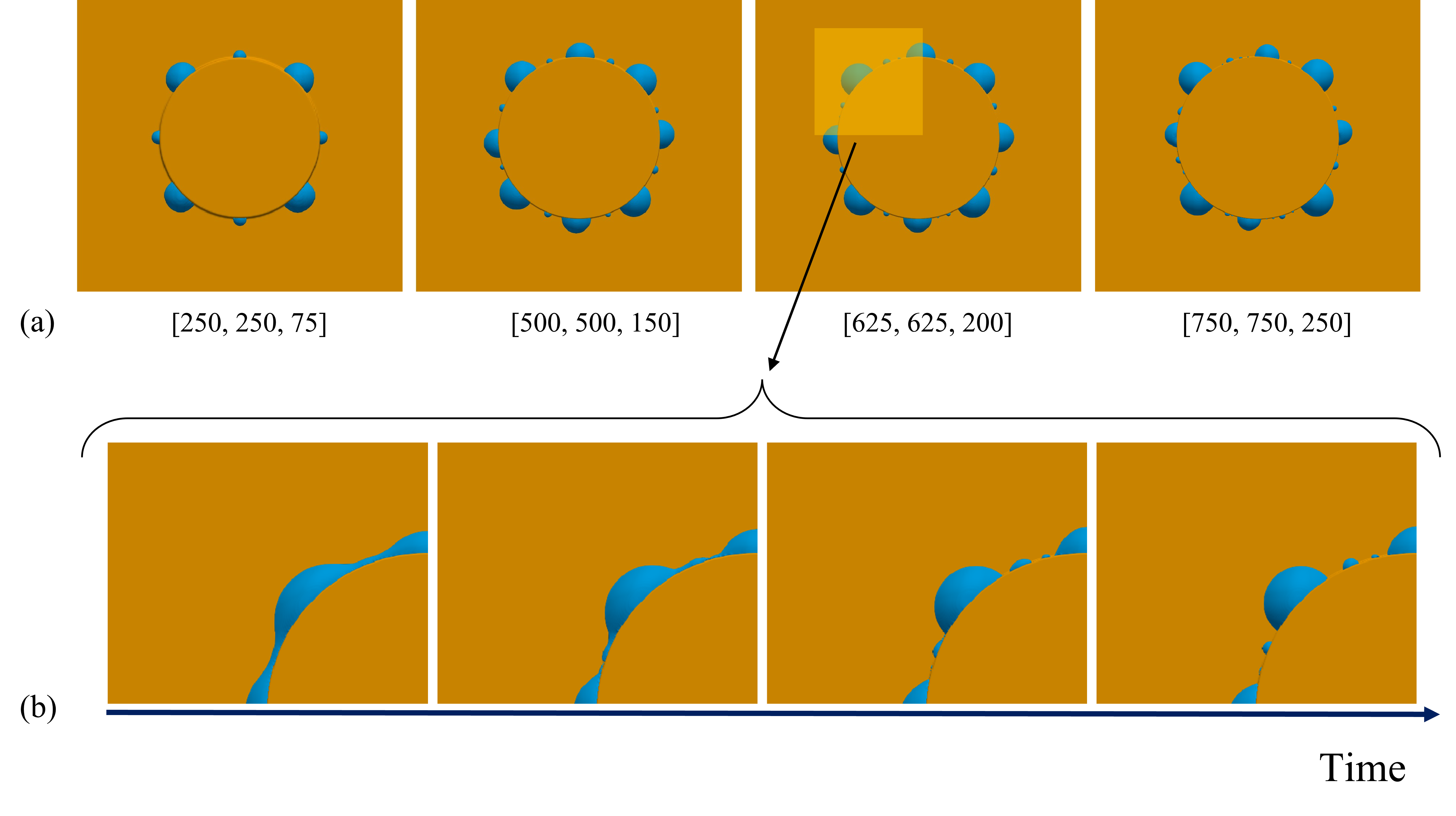}
  \captionsetup{justification=justified}
  \caption{Grid dependence study for the test case with $r_{w0}=0.13$, $\theta_1 = 90^{\circ}$ and $\theta_2 = 30^{\circ}$. (a) The post-instability patterns of four grid resolutions; (b) Zoom-in snapshots of the film breakup process for the grid resolution $[625, 625, 200]$.}
  \label{fig:gridStudy}
 \end{center}
\vspace*{6mm}
\end{minipage}

We perform a grid dependence study for the VOF simulation case with the thinnest wrapping film: $r_{w0}=0.13$, $\theta_1 = 90^{\circ}$ and $\theta_2 = 30^{\circ}$. Four different grid resolutions, $[250, 250, 75]$, $[500, 500, 150]$, $[625, 625, 200]$ and $[750, 750, 250]$, are investigated. All other physical and computational parameters are kept constant
for all cases. Figure \hyperref[fig:gridStudy]{\ref*{fig:gridStudy}(a)} shows the post-instability patterns of the four grid resolutions. Except the lowest grid resolution $[250, 250, 75]$, all other grid resolutions produce similar results, with the wrapping film eventually breaking up into eight satellite droplets. Notably, when a neck region approaches to be pinched off, as shown in figure \hyperref[fig:gridStudy]{\ref*{fig:gridStudy}(b)}, the film becomes extremely thin and its size is even close to the grid spacing. Thus, the VOF simulation can fail to capture the interfacial dynamics at the breakup stage. Therefore, the size and number of sub-droplets appearing after breakup strongly depend on the grid resolution. However, the emergence of these sub-droplets does not affect the major satellite droplets which stably form before breakup.

\FloatBarrier
\bibliographystyle{jfm}
\bibliography{jfm-instructions}
\end{document}